\documentclass{MScthesisITEM}

\usepackage{blindtext}
\usepackage{amsmath}
\usepackage{float}
\usepackage{listings}
\usepackage[utf8]{inputenc}
\usepackage[acronym]{glossaries}

\usepackage{blindtext}
\usepackage{amsmath}
\usepackage{float}
\usepackage{listings}
\usepackage{glossaries}

\usepackage{algorithm}
\usepackage{algorithmicx}

\title{Personalized Search} 
\author{Fredrik Nygård Carlsen\newlinetitle}

\supervisor{Trond Aalberg, IDI}

\DeclareGraphicsExtensions{.pdf,.jpg,.png,.jpeg}
\graphicspath{{./fig/}}

\newacronym{ntnu}{NTNU}{Norwegian University of Science and Technology}

\newacronym{prf}{PRF}{Pseudo relevance feedback}

\newacronym{irf}{IRF}{Implicit Relevance Feedback}

\newacronym{solr}{Solr}{Apache Solr}

\newglossaryentry{drank}{type=main, name={d-Rank}, description={Score aggregation method}}

\newacronym{marc}{MARC}{\textbf{MA}chine \textbf{R}eadable \textbf{C}ataloging}

\newacronym{oai-pmh}{OAi-PMH}{Open Archives Initiative - Protocol for Metadata Harvesting}

\newacronym{mlt}{MLT}{More Like This}

\newacronym{bfs}{BFS}{Breath-First Search}

\newacronym{qps}{QPS}{Queries per second}

\newacronym{ucd}{UCD}{User Centric Design}

\newacronym{originallist}{OL}{Original List}

\newacronym{vsm}{VSM}{Vector Space Model}

\newacronym{bm}{BM}{Boolean model}

\newacronym{cds}{CDS}{CERN Document Server}

\newacronym{cern}{CERN}{The European Organization for Nuclear Research}

\newacronym{tf-idf}{tf-idf}{term frequency - inverse document frequency (tf-idf)}

\newacronym{jif}{JIF}{Journal Impact Factor}

\newacronym{ir}{IR}{Information Retrieval}

\newacronym{cms}{CMS}{Content Management System}

\newacronym{mltdl}{MLTDL}{More Like This for Digital Libraries}

\newglossaryentry{Recommendation system}{type=main, name={Recommendation system}, description={A recommendation system is an automatic search that uses information such as user behaviour and personal information to predict interesting elements}}

\newglossaryentry{global search}{type=main, name={Global Search}, description={Searching all collections}}

\newglossaryentry{hitset}{type=main, name={Hitset}, description={The resulting document corpus when retrieving search results (often by applying a similarity model)}}

\newglossaryentry{invenio}{type=main, name={Invenio}, description={Digital Library system developed at CERN}}

\newglossaryentry{item}{type=main, name={Item}, description={A general term that includes documents as well as non-textual information, such as multimedia objects}}

\makeglossaries

\begin{document}
\selectlanguage{english}
\pagenumbering{roman}
\pagestyle{plain}

\titleITEM


\selectlanguage{english}
\renewcommand{\abstractname}{Abstract}
\begin{abstract}
As the volume of electronically available information grows, relevant items become harder to find. This work presents an approach to  personalizing search results in scientific publication databases. 
This work focuses on re-ranking search results from existing search engines like Solr or ElasticSearch. This work also includes the development of Obelix, a new recommendation system used to re-rank search results. The project was proposed and performed at CERN, using the scientific publications available on the CERN Document Server (CDS). This work experiments with re-ranking using offline and online evaluation of users and documents in CDS. The experiments conclude that the personalized search result outperform both latest first and word similarity in terms of click position in the search result for global search in CDS.

\end{abstract}

\clearpage

\addcontentsline{toc}{section}{Acknowledgment}

\renewcommand{\abstractname}{Acknowledgments}
\begin{abstract}

I would first of all like to thank my supervisor, Trond Aalberg, for valuable guidance and support. Furthermore, I would like to thank the people at CERN: Head Librarian at Invenio, Jens Vigen; CDS Developer, Patrick Glauner; Head of CDS, Jean-Yves Le Meur, and Head developer at Invenio, Tibor Simko, for valuable insight and ideas. I would like to thank my friends who supported me through the entire process; Thea Christine Mathisen and Rolf Erik Lekang.

\end{abstract}


\selectlanguage{english}

\renewcommand{\bf}{\textbf}

\tableofcontents*
\cleardoublepage

\listoffigures

\listoftables
\clearpage

\printglossary[title=Glossary, style=long]
\cleardoublepage
\glsaddall[]

\printglossary[title=List of Acronyms,type=\acronymtype] 

\pagenumbering{arabic}
\pagestyle{ruled}
\chapter{Introduction}

\section{Motivation}
Search engines have changed the way users find, interact with, and access information. Users can quickly search for and discover relevant information by issuing queries to search engines. Notwithstanding the growing popularity, search engines are facing a number of challenges. For example, given a query, a typical search engine returns a long list of ranked items, usually displayed over a number of pages, referred to in this work as the \gls{originallist}. However, the ranking algorithm implemented in most search engines is too simple and may not always satisfy the user's information needs. Users may have to scroll through several pages in order to find the desired information, affecting the user's search experience and overall satisfaction \cite{Vesely:1306230}.

In order to address the issue of irrelevant ranking, personalized search has attracted much interest in both academia \cite{gauch2007user} \cite{journals/fss/VictorCCS09} \cite{Vesely:1306230} \cite{krishnan2008predicts} \cite{Teevan:2005:PSV:1076034.1076111} \cite{qiufeng} \cite{bonett2001personalization} \cite{yoan} and in the enterprise world, for instance at Amazon \cite{Linden:2003:ARI:642462.642471}, Netflix \cite{Bennett07thenetflix}, Facebook \cite{Yue:2014:PCC:2566486.2567991}, Yahoo! \cite{sarukkai2013customizable}, Microsoft \cite{microsoftevaluation}, and Google \cite{brukman2013systems} \cite{culliss2003personalized}. Personalizing the search result based on an individual user's interests is made possible by the premise that a user may help the search engine disambiguate the actual intention of a query. However, studies have shown that the vast majority of users are reluctant to provide any explicit feedback on search results and their interests \cite{qiufeng} \cite{Vesely:1306230}. 

This work focuses on how search engines in digital libraries can learn a user's preference automatically, based on their search history, by building a graph representing how the users and items relate. Most of the research performed on personalized search focuses on general web search \cite{ilprints422} \cite{gauch2007user} \cite{bonett2001personalization} \cite{qiufeng} \cite{Linden:2003:ARI:642462.642471} \cite{microsoftevaluation} \cite{krishnan2008predicts} \cite{sarukkai2013customizable} \cite{brukman2013systems} \cite{ferragina2008personalized} \cite{journals/fss/VictorCCS09}, whereas the goal of this work is to improve search in digital libraries, using techniques and current knowledge of recommendation systems and personalization.

Related work by (see section \ref{subsec:related_work}) has shown that the users' preferences can be learned accurately even from small data sets of click history. The work by Qiu and Feng \cite{qiufeng} also demonstrates that personalized search based on user preference yields significant improvements compared to traditional ranking mechanisms.

The demand for better search engines has driven the development of personalization and recommendation systems, attempting to predict the users' interests and use this information to assist users in finding relevant items. Rather than relying on the users' unrealistic ability to write perfect queries to match their intent when searching, the recommendation systems use implicit or explicit feedback to build recommendations based on the users' interests. \cite{Teevan:2005:PSV:1076034.1076111}

Recent work on personalized collections by Blanc \cite{yoan} has shown a lacking search experience due to similar search phrases with different intentions. For example, a physicist may issue the query “Geneva” to look for information about CERN while a politician may issue the same query to find information about the United Nations or the Red Cross. Personalized search attempts to solve the disambiguate search results by including the users' preferences or intentions when ranking. 

This work explores models of user interests, built from their interaction with the digital library, including what the users search for as well as the usage of items.

\section{Problem Formulation}
Personalized search refers to search experiences that are specifically tailored to an individual's interests by incorporating information about the individual beyond the specific query provided. The topic of this thesis is to survey existing techniques for personalized search and propose, implement, test and evaluate a prototype for the CERN Document Server. 

\section{Context}

\begin{description}
\item[Available data sets] \hfill \\ This work utilizes data sets provided by CERN. The data sets include logs from the past ten years, as well as a database containing the current users and items in \gls{cds}.

\item[Content of data sets] \hfill \\ The logs are collected from the web servers, load balancers, and databases. The logs of interest are the ones from the load balancers, which aggregate all logs from the web servers, including all interactions between users and items, such as clicks, searches, views and downloads. These logs contain more than 500 million entries. Logs from other services like database is ignored in this work, as they do not provide any relevant information. The relevant logs contain information about the URLs visited and connects it with the user performing the actions.

\item[Unique users] \hfill \\ The logs contain user-IDs for logged-in users. For users that are not logged-in, only the used IP address is available, making it impossible to identify uniquely users correctly. Identifying the users that are not logged-in is challenging because multiple users may share the same IP address, and a user may change IP address over time. For that reason, only interactions by logged-in users are accounted for.

\item[Time constraints] \hfill \\ The duration of this project is fixed to last for five months, limiting the range and length of experiments.    

\item[Computing power] \hfill \\ This project is granted two virtual machines, each with 8 cores and 16 GB of ram, limiting tests of scalability over multiple machines.

\end{description}

\section{Approach}
The primary motivation for this work is based on feedback from real users of \gls{cds} and previous work performed at CERN on search and personalized experiences for CDS \cite{yoan} \cite{Vesely:1306230} \cite{marian2010} \cite{Gvianishvili:1295600}. 

Consequently, a reasonable starting point is to survey existing implemented techniques, previous research performed at CERN, and any work by other researchers in the field of personalized search and recommendation systems. This work explores the state of the art for ranking and personalization, as well as the current implementations at CERN and why it does not sufficiently satisfy the users' information needs.

This work includes implementing a new recommendation system (Obelix) that attempts to recommend relevant items for the users. The next chapters cover the process of developing Obelix and using this new recommendation system to personalize the search experience of CDS. \\

The approach is divided into five steps, as shown below. The \textit{prototype} refers to the entire system of personalized search, including Obelix and an integration with a digital library (CDS in the case of CERN).  

\begin{itemize}
  \item Survey existing techniques and procedures in the field of information retrieval (chapter 2).
  \item Survey related work, both from \acrshort{cern} and other sources (chapter 3). 
  \item Propose a prototype of personalized search for CERN Document Server (chapter 3).
  \item Implement the aforementioned prototype (chapter 4).
  \item Test and evaluate the aforementioned prototype with real users of CDS (chapter 5).
\end{itemize}




\chapter{Background}
\label{chap:background}
This chapter explores information retrieval, digital libraries (Invenio in particular), recommendation systems, personalized search, and evaluation methods for IR. Starting with an overview of fundamental concepts in Digital Libraries and \gls{ir}, followed up by more thorough theories on recommendation systems and personalization, and finishing up with evaluation methods.

\section{Digital Libraries}

\begin{quote}
A Digital Library: A possibly virtual organization that comprehensively collects, manages, and preserves for the long term rich digital content, and offers to its user communities specialized functionality on that content, of measurable quality and according to codified policies. (DELOS Manifesto) \cite{delos_manifesto} 
\end{quote}

A digital library is a quality controlled collection of items, managed and stored in a digital format (as opposed to physical books or DVDs). Such a library is often equipped with an interface for search and retrieval of specific items. Some well-known examples of digital libraries include Google Scholar, Google Books, and DiVA \cite{chowdhury2002introduction}.

A wide range of digital libraries exist, \textit{Invenio}, \textit{Fedora DSpace}, \textit{Fedora Commons}, \textit{EPrints}, \textit{Greenstone}, and \textit{Omeka} being the most referenced digital library systems in the literature.

\subsection{Invenio Digital Library}
\label{bg:invenio}
Invenio is an open source software package that provides the tools for management of digital assets in an institutional repository. It was developed by the \textit{CERN Document Server Software Consortium} and is freely available for download. Invenio complies with standards such as the Open Archives Initiative metadata harvesting protocol (OAI-PMH) and uses MARC 21 as its underlying bibliographic format. The technology offered by the software covers all aspects of digital library management from item ingestion, through classification, indexing, and curation, to dissemination \cite{Canevet:1479357}. Invenio was originally developed at CERN to run the CERN Document Server (CDS), but is currently being co-developed by an international collaboration comprising institutes (CERN, DESY, EPFL, FNAL, and SLAC).

Besides being used to run the CERN Document Server Server, Invenio has also been chosen by several other major institutions and projects, Invenio is currently in use by approximately 30 scientific institutions worldwide. Among them is the INSPIRE service, the reference repository for High Energy Physics items \cite{Praczyk:1624334}. As Invenio is a research project, the focus on modularity is apparent in the architecture. It consists of several more or less independent modules with their own responsibility. 

\begin{figure}[h!]
  \centering
    \includegraphics[width=1\textwidth]{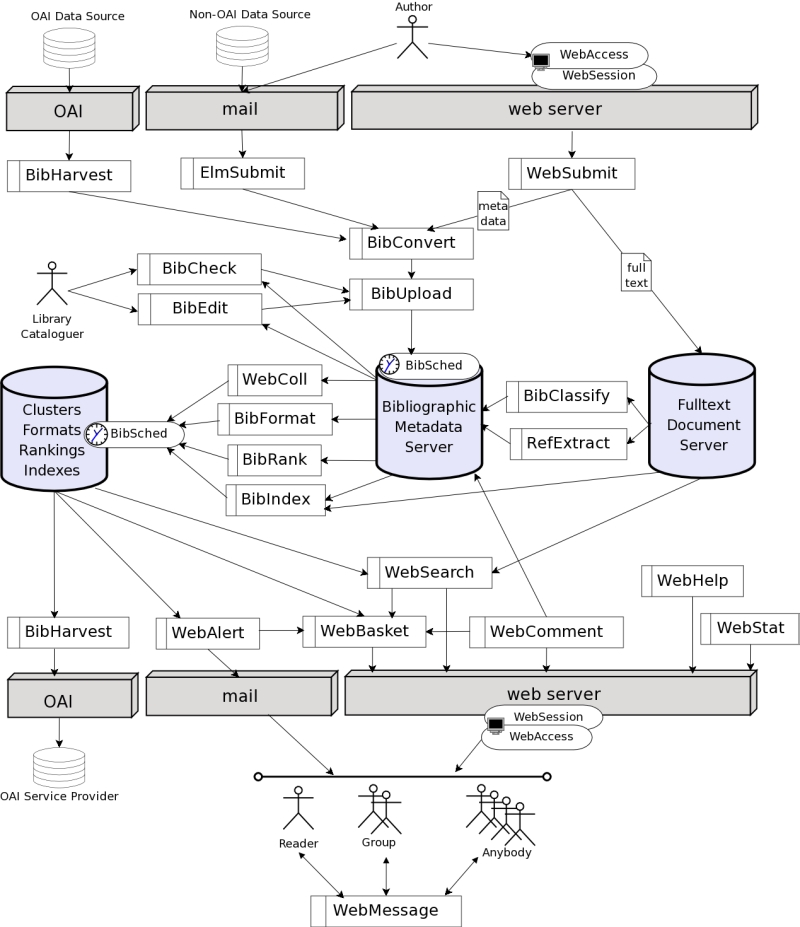}
  \caption{Invenio Modules}
  \label{fig:invenio_module_relationship}
\end{figure}

To provide a basic overview of Invenio's architecture and structure, the cornerstone modules are presented below. A complete overview of Invenio's modules can be found in appendix \ref{App:AppendixA}. A diagram of the relationship between the different modules can be found in figure \ref{fig:invenio_module_relationship}. The modules use the Bib prefix to indicate a focus on the bibliographic data, and the Web prefix to indicate focus on the web interface. In some cases, the focus is more blurred (e.g. the search engine which is used through the web interface, but works on bibliographic data) \cite{Gvianishvili:1295600}.

\subsubsection*{Cornerstone modules of Invenio}

\begin{itemize}
    \item \textbf{BibConvert, BibCheck, BibClassify}. These modules play an important role in automating the item import. BibConvert is a powerful tool that converts metadata from hundreds of sources into the MARC XML format used in Invenio. BibCheck makes sure the metadata is coherent with quality standards, and BibClassify classifies the item using a controlled vocabulary.
    
    \item \textbf{OAiHarvest}. Responsible for automatic harvesting of OAi PMH-compliant repositories. This enables tight integration with other digital libraries, and allows for global cooperation of library services.

    \item \textbf{BibIndex}. Invenio's indexing engine. Consists of a forward index listing, with words/phrases as the key and location records as values; and a reverse index listing, where each record is registered with a set of words/phrases.
    
    \item \textbf{BibRank}. Uses term frequency, freshness, popularity and other parameters to create a ranking for each record, for use in the search engine.
    
    \item \textbf{WebAccess}. The module responsible enabling web access for users to maintain the different aspects of the system. By using a Role-Based Access Control, users are allowed administrative access based on their role (such as «librarian» or «system developer»).

    \item \textbf{WebSearch}. Invenio's search module is backed by several indices, and is designed to provide fast responses. The metadata is also directly browsable.
\end{itemize}

\noindent{
In addition to the ones listed above, there are several modules which mainly adds usability, such as WebComment, WebMessage and WebBasket, which constitutes the heart of the social network features of Invenio. A full overview is available in appendix \ref{App:AppendixA}.
}

\section{Information Retrieval}
\begin{quote}
Information Retrieval (IR) is the activity of obtaining information resources relevant to an information need from a collection of information resources \cite{Belkin92informationfiltering}. 
\end{quote}

The goal of \gls{ir} systems is to provide users with items that will satisfy their information need \cite{Xu:1996:QEU:243199.243202}. The word \textit{item} is used as a general term that includes documents as well as non-textual information, such as multimedia objects. 

\gls{ir} models define the procedures for how a query executes and generates search results. The goal of every retrieval model is to produce search results with high recall and high precision \cite{microsoftevaluation}. In simple terms, high precision indicates that an algorithm returned substantially more relevant results than irrelevant, whereas high recall indicates that an algorithm returned most of the relevant results (more on precision and recall in section \ref{subsub:measuring_usage_prediction}).

The following major models have been developed to retrieve information: the Boolean model, the Statistical model (which includes the vector space and the probabilistic retrieval model). The first model is often referred to as the \textit{exact match model}, the latter ones as the \textit{best match models} \cite{Belkin92informationfiltering}. The following subsections provide an overview of the fundamental information retrieval models relevant for this research. 

The information retrieval process may be split into two tasks, the retrieval of items and the ranking of the retrieved items. The retrieval is often performed using an inverted index, which contains of all the indexed terms. When a user type in a query, the query is matched against the inverted index, returning all items that match at least one of the terms in the query. Ranking methods are further used to decide which items are most relevant, and based on this information decides the ordering of the items in the search result.

\subsection{Inverted Index}
The inverted index is the central data structure in most information retrieval systems \cite{Manning:2008:IIR:1394399}. The simplest implementation provides a mapping between terms and their location of occurrence in the collection of items. The inverted index consists of a \textit{dictionary}. Each searchable term has an associated \textit{postings list}, a list with all items where that term occur. The postings list may be implemented as a list of items or a list of positions where the term appears within the items. 

The inverted index is often used to retrieve \textit{matching} items for a given query \cite{Manning:2008:IIR:1394399}. When a user performs a search, the search engine may return all items from the corresponding postings list for the terms used in the query. Those items are then further ranked using a chosen ranking method.

\subsection{Retrieval and Ranking Methods}
Building on the data structure of the inverted index, this section presents retrieval methods to rank and retrieve items. Two popular techniques are presented first, the Boolean model and Vector Space Model (VSM), which is heavily used in popular search engines like Solr or ElasticSearch. The third technique is the \textit{h-index}, a ranking method using citation counts. 

\subsubsection{The Boolean Model}
\label{subsec:bgBooleanmodel}
The Boolean model is a simple similarity model based on Boolean algebra. As the name suggests, queries are considered from a \textit{true/false} point of view, where the items either match the query, or they do not. The queries used with the Boolean model are expressions of Boolean algebra, where each variable is a query term. An algorithm implementing a Boolean model will return matching items without any further ranking. For example, the query [q = $t_1$ OR ($t_2$ AND $t_3$)] would return any item that either contains term $t_1$, or both $t_2$ and $t_3$ (or all three terms).

\subsubsection{Vector Space Model}
\label{subsec:bgvectorspacemodel}

The vector space model seeks to extend the limitations imposed by the Boolean model and introduces a way of partial matching. It represents items as \emph{n}-dimensional vectors, where \emph{n} is the number of indexed terms in the \gls{ir} system. The cosine similarity model creates a score for each item by calculating the scalar product between each item vector and the query vector. The score is correlated with similarity, a high score indicates high similarity to the query vector \cite{Manning:2008:IIR:1394399}.

$$
sim(q,d) \; = \; \vec{v(q)} . \vec{v(d)} \; = \;
\frac{\vec{V(q)} . \vec{V(d)}}{\| \vec{V(q)} \| . \|
  \vec{V(d)} \|}
$$

\subsubsection{TF-IDF}
\label{subsec:tf-idf}
Tf-idf stands for term frequency-inverse document frequency, and the tf-idf weight is a weight often used in information retrieval and text mining. This weight is a statistical measure used to evaluate how important a word is to a document in a collection or corpus. The importance increases proportionally to the number of times a word appears in the document but is offset by the frequency of the word in the corpus. Variations of the tf-idf weighting scheme are often used by search engines as a central tool in scoring and ranking a document's relevance given a user query \cite{Manning:2008:IIR:1394399}.

\begin{description}
\item[TF] \hfill \\
Term Frequency, which measures how frequently a term occurs in a document. Since every document is different in length, it is possible that a term would appear much more times in long documents than shorter ones. Thus, the term frequency is often divided by the document length \cite{Manning:2008:IIR:1394399}. 
\item[IDF] \hfill \\
Inverse Document Frequency, which measures how important a term is. While computing TF, all terms are considered equally important. However it is known that certain terms, such as \textit{is}, \textit{of}, and \textit{that}, may appear often without importance. Thus we need to weigh down the frequent terms while scale up the rare ones \cite{Manning:2008:IIR:1394399}.
\end{description}

Term weighting using the $tf-idf_{t,d}$ measure 
$$
tf-idf_{t,d} \; = \; tf_{t,d} \times idf_t \; = \; tf_{t,d}
\times log(\frac{N}{df_t})
$$

\subsection{h-index}
\label{sec:bghindex}
The h-index is based on the distribution of citations received by a given researcher's publications. Hirsch writes: 
\begin{quote}
A scientist has index $h$ if $h$ of his/her $N_p$ papers have at least $h$ citations each, and the other ($N_p - h$) papers have no more than $h$ citations each \cite{Hirsch:2010:IQI:1873250.1873262}. 
\end{quote}
Hirsch proposes the index $h$, defined as the number of papers with citation number $>h$, as a useful index to characterize the scientific output of a researcher.

\begin{figure}
  \caption{h-index}
  \label{fig:hindex_graph}
  \centering
    \includegraphics[width=0.5\textwidth]{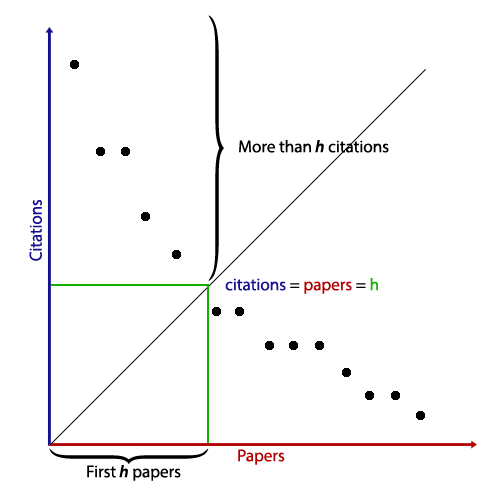}
\end{figure}

As illustrated in figure \ref{fig:hindex_graph}, the h-index plot shows correlation between  publications and the number of citations per publication. The h-index is designed to give an improved measurement upon simpler methods such as count of citations divided by total count of papers. The index works for scientists working in the same field, but may be misleading across different research communities \cite{Hirsch:2010:IQI:1873250.1873262}. 

\section{Recommendation Systems}
Recommendation systems have become common in recent years and are a subclass of information filtering system that seek to predict the rating or preference that user would give to an item \cite{DBLP:reference/rsh/2011} \cite{howcomputersworks}.

The most popular applications of recommendation systems are probably movies, music, news, books, research articles, search queries, social tags, and products in general. However, there are also recommendation systems for such as restaurants \cite{sawantyelp}, financial services \cite{paper:felfernig:2007} and Twitter followers \cite{Gupta:2013:WFS:2488388.2488433}. \\

\begin{quote}
A recommendation system uses user preferences, gathered implicitly or explicitly, to generate a personalized list of recommended items for a user - through collaborative or content-based filtering \cite{leino2014user}.
\end{quote}

Multiple definitions of recommendation systems exist, the chosen definition emphasizes the need for recommendation systems. The increasing number of items and information to consider requires a technological tool to handle the information overload and to help users find relevant information \cite{leino2014user}. 

Recommendation systems can be an efficient technique to meet the challenge of information overload, as they help narrow down the items to consider by suggesting items we are likely to be interested in as well as ordering the search result by relevance for a given user. In effect, they help users discover relevant items quicker, finding items they would not have found otherwise using traditional text matching techniques such as TF-IDF, using Solr or equivalent systems. Existing research on recommendation systems has led to a widely adoption by e-commerce (Amazon \cite{Linden:2003:ARI:642462.642471}) and other systems (Netflix \cite{Bennett07thenetflix}) where the number of items exceeds the ability to find the relevant items efficiently. 

There exist two main approaches to recommendation systems, collaborative filtering and content-based filtering. In addition, a hybrid approach which combine the two can be more effective in some cases. For instance, Netflix is such a system which combines the searching habits of similar users (i.e. collaborative filtering) as well as offering movies that share characteristics with films that a user has given a high rating (content-based filtering).

\subsection{Collaborative Filtering}
\label{sec:background:collaborative_filtering}
There are two main types of collaborative filtering algorithms in the recommendation system literature, memory-based and model-based. 

\begin{description}

\item[Memory-based] \hfill \\ 
Memory-based collaborative filtering utilizes the entire data set to generate predictions. Using statistical techniques to find users or items that have been similar in the past, either a group of users who rates the same items or items that were all rated the same way by the same users. Then some algorithm takes advantage of these similarities to build predictions and recommendations for users \cite{sarwar2001item}.

\item[Model-based] \hfill \\ 
Model-based collaborative filtering algorithms provide item recommendation by first developing a model of user ratings. Algorithms in this category take a probabilistic approach and envision the collaborative filtering process as computing the expected value of a user prediction, given his/her ratings on other items. The model building process is performed by different machine learning algorithms such as Bayesian network, clustering, and rule-based approaches \cite{sarwar2001item}. 

\end{description}

\subsection{Content-Based Filtering}
Content-based filtering, also referred to as \textit{cognitive filtering}, recommends items based on a comparison between the content of the items and a user profile. The content of each item is represented by a set of descriptors or terms, typically the words that occur in a item. The user profile is represented by the same terms and built up by analyzing the content of items that have been seen by the user \cite{sarwar2001item}.

A key issue with content-based filtering is whether the system is able to learn user preferences from user's actions regarding one content source and use them across other content types. When the system is limited to recommending content of the same type as the user is already using, the value from the recommendation system is significantly less than when other content types from other services can be recommended. For example, recommending news articles based on browsing of news is useful, but it's much more useful when music, videos, products, discussions etc. from different services can be recommended based on news browsing \cite{opac-b1122543}.

\subsection{Trust-enhanced recommendation systems}
\label{background:trustenhanced}
Recommendations generated based on information coming from trust networks, a social network that express how many users of the community trust each other. Users receive recommendations for items recommended by users they trust or users who are trusted by users you trust. The main strength of systems like this is the use of trust propagation and trust aggregation, the mechanism to calculate the transitivity of users' trusts throughout the system \cite{Heb:2008:TRM:1816870}. There are two main approaches to trust-enhanced recommendation systems, a probablistic and a gradual approach.

\begin{itemize}
\item A probabilistic approach deals with binary values for trust. A resource may be trusted or distrusted, and computes a probability that the resource can be trusted. The system sets a threshold value required to be trusted. The trust between two users in such a system can be computed based on how many trusted connections there are between the two users. (A connection between two users is often named a transaction.) The score is a sum of transactions divided by the number transactions. \cite{Heb:2008:TRM:1816870} 
\item A gradual approach also takes into account to what extent the users trust each other; the user may trust another user on a scale from 1 to 5 or some other scale suitable for the data set. In a gradual approach, trust values are not interpreted as probabilities, a higher value corresponds to a higher trust, not whether a user trust another user or not, as in the probabilistic approach. \cite{Heb:2008:TRM:1816870} 
\end{itemize}

Trust networks among users of a recommendation system prove beneficial to the quality and amount of the recommendations \cite{journals/fss/VictorCCS09}. Trust networks allow users to establish better-informed opinions about items through the judgement of other users they trust \cite{Heb:2008:TRM:1816870}. Trust networks can contribute to the success of recommendation systems by allowing users to establish better-informed opinions about certain items through the judgment of trusted sources/agents that have evaluated or experienced those items. Trusted agents can make additional recommendations over the ones generated by other RS techniques, which especially benefit users who lack a properly detailed user profile. 

The transitivity is computed based on how much user $a$ trusts user $c$, given the value of of trust user $a$ has in user $b$ and user \textit{b} has in user \textit{c} (propagation) and by combining several trust values into one final trust value (aggregation). Propagation and aggregation are the key elements of building \textit{trust metrics}, which aim to estimate trust between two users in a social recommendation system \cite{oreilly2005design}.

\subsubsection{Distrust}

In a large group of users, it is only natural to also find \textit{distrust}; some users you trust to some degree, but some users you simply do not trust. Most approaches completely ignore distrust \cite{Heb:2008:TRM:1816870} or simply consider distrust as a low value on the trust scale. However, there is a growing opinion that distrust cannot be seen as the equivalent of lack of trust \cite{Heb:2008:TRM:1816870}.

\begin{figure}
  \centering
    \includegraphics[width=0.5\textwidth]{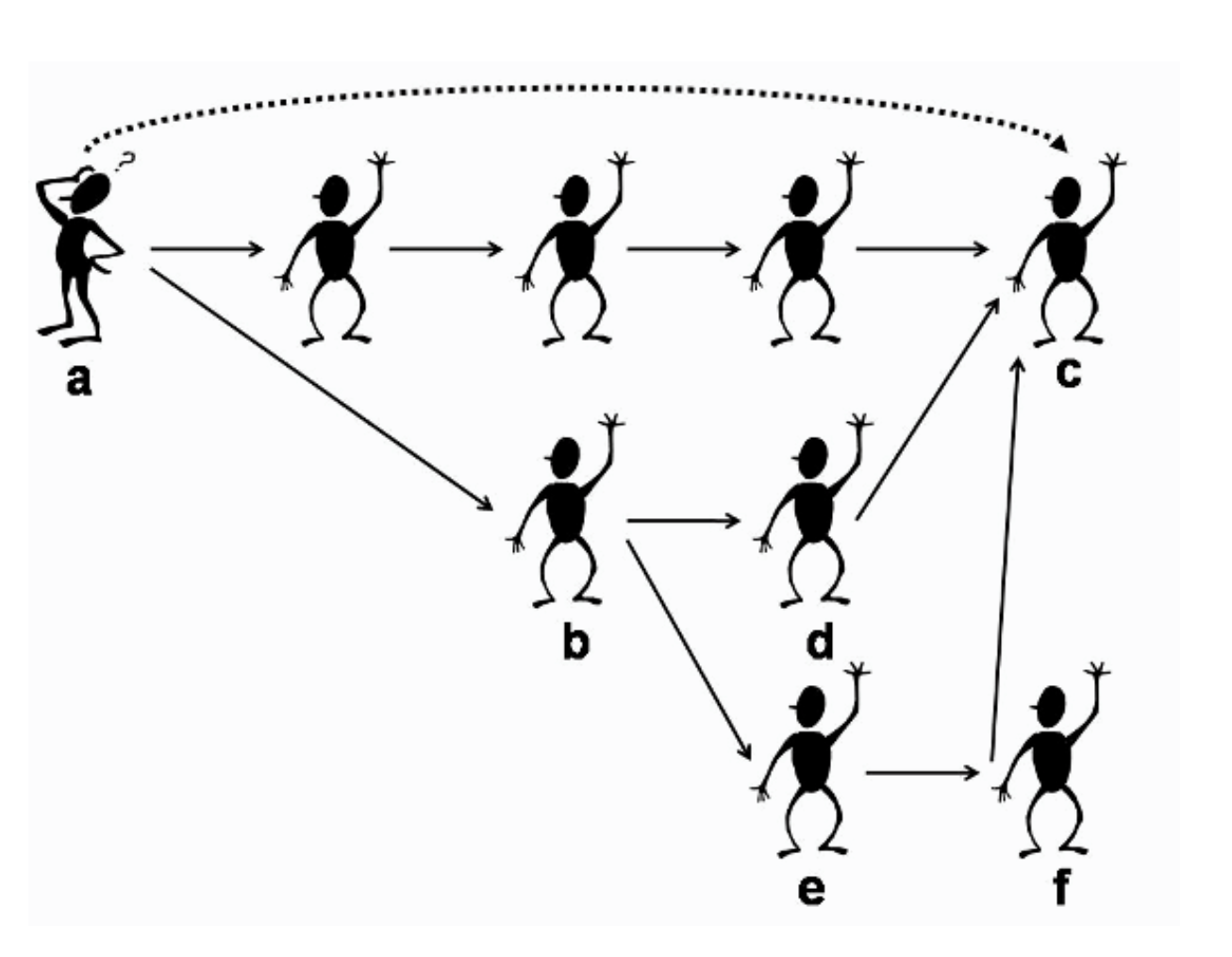}
  \caption{Trust Network}
\end{figure}

In the case of atomic direct propagation, if $a$ trusts $b$ and $b$ trusts $c$, $a$ might trust $c$ to a certain degree. Analogously, if $a$ trusts $b$ and $b$ distrusts $c$, it seems clear that $a$ should somehow distrust $c$. However, the picture gets more complicated when we also allow distrust as the first link in a propagation chain. For example, if $a$ distrusts $b$ and $b$ distrusts $c$, there are several options for the trust estimation of $a$ in $c$: $a$ possible reaction is to infer that $a$ should trust $c$, since $a$ might think that distrusted acquaintances of users he distrusts are best to be trusted (‘the enemy of your enemy is your friend’). Or $a$ should distrust $c$ because $a$ thinks that someone that is distrusted by a user that he distrusts certainly must be distrusted. Yet another interpretation of distrust propagation is to ignore information coming from a distrusted user $b$, because $a$ might decide not to take into account anything that a distrusted user says \cite{Heb:2008:TRM:1816870}.

\subsection{Benefits of Recommendation Systems}
Recommendation systems may benefit both the users and service providers. Recommendation systems are at the center of personalization, effectively it is impossible for two users to have exactly the same experience using Amazon.com during any longer visit to the site. 

The personalization begins at the moment the user enters the website (using cookies, location, etc) and continues to develop as the user explores the content of the website \cite{Linden:2003:ARI:642462.642471}.  

\subsection{User Relevance Feedback (URF)}
\label{sec:relevance-feedback}
The purpose of user relevance feedback is to involve the user in the retrieval process. The user gives either explicit or implicit feedback on the relevance of retrieved items. The feedback is gathered and used to improve search results. 

Explicit feedback is a method where the user of a system gives input, either by rearranging the results from a query or click on some predefined button that says \textit{this is relevant to my query}. The feedback is defined as explicit only when the user knows that the feedback provided is treated as relevance judgments. The user can indicate the relevance either as a binary value, (relevant, not relevant), or by using a graded system like (not relevant, somewhat relevant or very relevant).

In contrast to the explicit feedback where the user knows that their feedback is treated as relevance judgments, the implicit feedback is given by the users interaction with the search engine \cite{Xu:1996:QEU:243199.243202}. For instance, which items the user does and does not select for viewing, the duration of time spent viewing a item, or interactions like scrolling or browsing actions. 

\newpage
\section{Personalization}
\begin{quote}
Personalization involves a process of gathering user-information during interaction with the user, which is then used to deliver appropriate content and services, tailor-made to the user's needs. The aim is to improve the user's experience of a service \cite{bonett2001personalization}.
\end{quote}

Personalization may refer to either automatic personalization based on collected data or manually based on the user's preference. Personalized search can improve precision in search results as well as improving the user experience. 

Personalization is a broad concept covering several fields. The key for personalization is to use available data to improve the user experience, we wish to improve the search experience by personalizing the results with relevance for the user. An important precision is that the personalization is implicit, where the opposite is explicit personalization, also known as customization. Customization includes both user settings and preferences.

For levels of personalization are defined in the litterature \cite{yoan}.

\begin{description}

\item[Generic] \hfill \\
All users receive the same recommendations. This is quintessentially a non-personal level of recommendation \cite{schafer2006recommender}.

\item[Demographic] \hfill \\
All members of the target group receive the same recommendations \cite{konstan2008introduction}. 

\item[Ephemeral] \hfill \\
Recommendations match the current activity of the user; recommendations respond to the user’s navigation and item selection. An example of this is Amazon.com’s \textit{customers who bought this also bought...} recommendations. \cite{krishnan2008predicts} 

\item[Persistent] \hfill \\
Recommendations match long-term interests of the user \cite{konstan2008introduction}. Persistence requires the user(s) to maintain persistent identities in the system, but also rewards them with mostly personalized recommendations \cite{sarwar2001item}. 

\end{description}

\section{Evaluation Methods}

\subsection{Prediction Accuracy}
\label{bg:prediction_accuracy}

The most discussed property in the recommendation system literature is prediction accuracy \cite{microsoftevaluation}. In the core of the vast majority of recommendation systems lies a prediction engine, predicting user opinions over items (rating of books) or the probability of usage (e.g. views). A fundamental assumption for any recommendation system is that a system that provides better predictions will be preferred by the user, motivating researchers to find algorithms that provide better predictions. \cite{microsoftevaluation} 

Three broad classes of prediction accuracy exist \cite{microsoftevaluation}, measuring the accuracy of rating predictions, usage predictions and ranking of items. 

\subsubsection{Measuring Ratings Prediction Accuracy}
For some applications, it is interesting to predict the rating a user would give to an item. For example, given a service like Netflix, the users rate the items (e.g. 1-star through 5-stars). 

Root Mean Squared Error (RMSE) is probably the most commonly used metric for prediction evaluations.  RMSE measures the difference between the value predicted by a model or an estimator, and the values observed during the test. The system generates predicted ratings $\bar{r}_i$ for a test $N$ of user-item pairs, then these values are compared to the known true ratings. \cite{DBLP:reference/rsh/2011} Typically, the true value of $r_i$ is obtained through user studies or online experiments. The RMSE between the predicted and actual rating is given by: 

\begin{equation}
RMSE = {\sqrt {\frac{1} {N}{\sum\limits_{i = 1}^N {(\bar{r}_{i} - r_{i} } })^{2} } }
\end{equation}

Mean Absolute Error (MAE) is a popular alternative, given by: 

\begin{equation}
MAE =  {\frac{1} {N}{\sum\limits_{i = 1}^N {|\bar{r}_{i} - r_{i}}}| }\end{equation}

Comparing MAE to RMSE, the RMSE disproportionably penalizes large errors. The effect is that MAE prefers a system that makes an error of 3 on one rating and 0 on three ratings, whereas RMSE prefers a system that makes an error of 2 on three ratings and 0 on the fourth.

\subsubsection{Measuring Usage Prediction}
\label{subsub:measuring_usage_prediction}
In many systems, the users do not rate items. However, they may still use (click, view or download) items, making predictions of usage interesting. For example, when a user starts to watch a movie on Netflix, the system suggests a new set of movies, given the movie just started. In this case, what is interesting is not the predicted rating of the movie by the user, but rather which movies that may also be interesting, given the movie started.  

When performing offline evaluation of usage prediction, the test data typically consist of items each user has used. The procedure is then to select a user, hide some of the users' selections and ask the system to predict a set of items that the user will use. This test will result in four possible outcomes, as shown in table \ref{table_posneg}

\begin{table}[h]
\centering
\begin{tabular}{|l|l|l|}
\hline
 & Recommended & Not recommended \\ \hline
Used & True Positive & False Negative \\ \hline
Not used & False Positive & True Negative \\ \hline
\end{tabular}
\caption{Possible outcomes of usage prediction}
\label{table_posneg}
\end{table}

Typically, the data is collected before the recommendation system is enabled, enforcing the assumption that even when recommended, unused items are uninteresting or useless to the user. This assumption may be false, for example, a user may not select an item because the user is not aware of the existence of the item. But after the introduction of the recommendation system, the user may select it, in this case, affecting the number of false positives.

\subsubsection{Ranking Measures}
\label{ranking_measures}
Some applications, for instance search applications, typically present a list of items to the user, either as a vertical or horizontal list, imposing a certain natural browsing order. For example in \gls{cds}, where the search results are listed in a table and labeled from 1 to 10 by default, indicating that result number 1 is a better hit than result number 2. These lists may be long and in \gls{cds}, the user may need to click on \textit{next page} to find more results. In this case, it is not interesting to predict an explicit rating or selecting a set of recommended items, as in the previous sections. Instead, what is interesting is the ordering of items according to the user's preferences, known as ranking items. Two approaches exists for performing the offline measurement of ranking, "Reference Ranking" and "Utility-based ranking". 

\begin{description}

\item[Using a Reference Ranking] \hfill \\
The reference ranking refers to a correct order of ranking. In order to evaluate a ranking with respect to a correct order, such a reference ranking is necessary to obtain. In the case where rating is available, it is possible to use the ratings with ties, showing the items in decreasing order, with items given the same ratings tied \cite{microsoftevaluation}. For example, in the case of Netflix, the movies given an average rating of five is on the top of the list, followed by the movies with an average rating of four. But in the case of \gls{cds}, ratings are not available. An alternative approach when only usage data is available may be to rank used items above unused items. However, this approach is only valid if the user was aware of all the unused items, but chose to not use them. \cite{microsoftevaluation} This data is possible to obtain by observing and distinguish which items the user has "seen" and which items the user has "used". The items seen may be results that the user has ignored from a search result (scrolled through, but not used) \cite{DBLP:reference/rsh/2011}. 

Both approaches of references ranking described above (rating and usage) tie items when there is no information about the user's relative preferences. However, recommendation systems for applications like the \gls{cds} need to rank items with no ties, in order to create a valuable search result. For this reason, the evaluation of ranking should not penalize the ordering of ranked items when tied. For example, the recommendation engine may return item $x$ before item $y$, but the order of $x$ and $y$ does not matter for the evaluation, as long as they have the same score (tied).

\item[Utility-based ranking] \hfill \\
A popular alternative to the reference ranking is the utility-based ranking, assuming that a list of recommendations is additive, given by the sum of the utilities of the individual recommendations. The reference ranking scores a ranking on its correlation with some "true" ranking. The utility-based ranking scores the ranking on the utility of the recommended item discounted by a factor that depends on its position in the list of recommendations. For example, one utility may be the likelihood of a user to observe an item at position $i$ in the search result. Typically, the user scans the search result from top to bottom, making it more likely to observe the first items, compared to the latest items \cite{DBLP:reference/rsh/2011}. 

Another interpretation of the discount is the probability that a user would observe a recommendation in a particular position in the list. The utility is then given by observations depending only on the item recommended.  Under this interpretation, the probability that a particular position in the recommendation list is observed is assumed to depend only on the position and not on the items that are recommended \cite{DBLP:reference/rsh/2011}. 

\end{description}

\subsection{Multivariate Testing}
\label{multivariate_testing}
Multivariate statistics is a form of statistics encompassing the simultaneous observation and analysis of more than one outcome variable. The application of multivariate statistics is multivariate analysis \cite{microsoftevaluation}. 

Multivariate statistics concerns understanding the different aims and background of each of the different forms of multivariate analysis, and how they relate to each other. The practical implementation of multivariate statistics to a particular problem may involve several types of univariate and multivariate analyzes in order to understand the relationships between variables and their relevance to the actual problem being studied \cite{microsoftevaluation}.

Multivariate testing is an extended version of A/B-testing where more than one variable is tested, which is necessary in order to compare a set of variables.

\section{Scalability}
\label{background_scaling}
Scalability is the ability to handle a growing amount of data and computation complexity. Application scalability refers to the improved performance of running applications on a scaled-up version of the system.  \cite{El-Rewini:2005:ACA:1044920}

\begin{itemize}
\item To scale horizontally (or scale out) means to add more nodes to a system, such as adding a new computer to a distributed software application. An example might involve scaling out from one Web server system to three. As computer prices have dropped and performance continues to increase, high-performance computing applications such as seismic analysis and biotechnology workloads have adopted low-cost "commodity" systems for tasks that once would have required supercomputers.  \cite{El-Rewini:2005:ACA:1044920}

\item To scale vertically (or scale up) means to add resources to a single node in a system, typically involving the addition of CPUs or memory to a single computer. Such vertical scaling of existing systems also enables them to use virtualization technology more effectively, as it provides more resources for the hosted set of operating system and application modules to share. Taking advantage of such resources can also be called "scaling up", such as expanding the number of Apache daemon processes currently running.  \cite{El-Rewini:2005:ACA:1044920}
 
\end{itemize}

\chapter{Search on CERN Document Server}
This chapter presents the state of the art of search on CDS, beginning with an introduction to CERN and the CERN Document Server Server. Then an overview of the current implementation and related work is presented, followed by an analysis of needs for the users of CDS. 

\section{CERN}
The European Organization for Nuclear Research (CERN) is the world’s largest particle physics laboratory, an international organization established in 1954 \cite{Annual:1710503}. The organization is based in the northwest suburbs of Geneva on the Franco–Swiss border. The term CERN is also used to refer to the laboratory, which employs just under 2,400 full-time employees, 1,500 part-time employees, and hosts some 10,000 visiting scientists and engineers, representing 608 universities and research facilities, and 113 nationalities \cite{Annual:1710503}. CERN’s main function is to provide the particle accelerators and other infrastructure needed for high-energy physics research. As a result, numerous experiments have been constructed at CERN following international collaborations. It is also the birthplace of the World Wide Web \cite{Annual:1710503}. The main site at Meyrin has a large computer center containing powerful data-processing facilities, primarily for experimental data analysis. Because of the need to make these facilities available to researchers elsewhere, it has historically been a major wide area networking hub \cite{Annual:1710503}.

\subsection{CDS - CERN Document Server Server}
\gls{cds} is an instance of Invenio \ref{bg:invenio} running at CERN. The creation of \gls{cds} is the original motivation for the development of Invenio. The digital library (\gls{cds}) contains over 1.4 million items divided into more than 500 collections such as articles, books, journals, photos, videos, and more. The number of unique users in \gls{cds} is more than 40 000, making CDS one of the largest digital libraries in the physics domain. It is ranked at the third largest digital library in the world \cite{ranking_cds_web} \cite{Gvianishvili:1295600} \cite{Pepe:853565}.

Although this work aims to focus on Personalized Search for digital libraries in general, CDS, with its items, users, and needs will form the basis for the proposed system. 

\section{State of the Art}
The section presents the state of the art of search in Invenio and present related work in the field of ranking and personalized search in digital libraries.

\subsection{CDS Search Engine}
The search engine in CDS supports a number of different query types, and allow the user to search through all items or specific collections. The results are shown as one list or clusterd by collections, based on what the user specify when searching. 

\begin{figure}[H]
  \centering
    \includegraphics[width=1\textwidth]{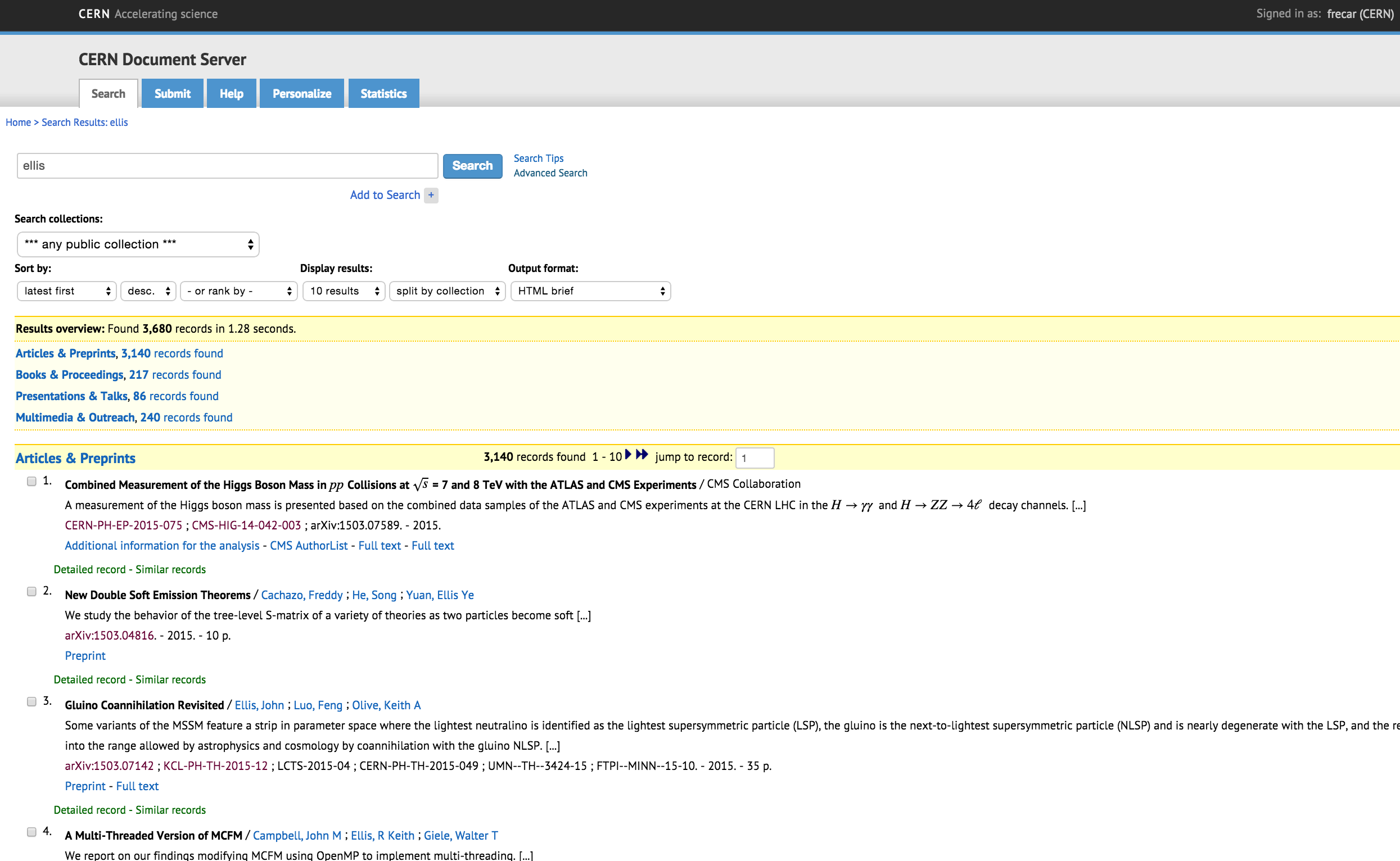}
  \caption{Example search for \textit{ellis} on CDS}
  \label{fig:example_search_cds}
\end{figure}

The search interface, as seen in figure \ref{fig:example_search_cds}, allows the user to choose between the simple search and advanced search. The simple search is similar to any other search engine like Google, whereas the advanced search allows the user to provide regular expressions and to search for terms in specific fields like title or description. The advanced search allows specifying fields, creating queries like \textit{title:Higgs AND abstract:CERN} or \textit{title:particle AND year:2015}. The current status of the implemented search engine may be summarized as follows:

\begin{description}
\item[Users] \hfill \\ The users registered at CERN uses a centralized authentication system, however, other users are also able to use CDS without login (most publications are available to the public).  

\item[Search] \hfill \\ The current search engine of CDS is solely based on metadata. When a user performs a search, the query is split into separate words and matched against the items in \gls{cds}. Additionally, there exists a feature that tries to guess what the user intended to search for when the query contains spelling mistakes.

\item[Full-text] \hfill \\ Full-text search is disabled due to the complex collections and item corpus in CDS, most publications mention the same words and are therefore difficult to distinguish between \cite{Vesely:1306230}.

\end{description}

\subsection{Related Work}
\label{subsec:related_work}

The continuous development of the search engine in CDS has resulted in a number of projects attempting to improve the search experience  \cite{vesely2012selection} \cite{Vesely:1306230} \cite{yoan} \cite{marian2010} \cite{Canevet:1479357}. 

The latest contribution, the \gls{drank} project \cite{vesely2012selection} \cite{Vesely:1306230} is a significant contribution to CDS, attempting to aggregate multiple ranking methods such as \textit{latest first}, \textit{citations}, \textit{tf-idf}, and \textit{download history}, in order to improve the ranking of search results. Other notable contributions to the Invenio search engine are the implementation of Solr by Patrick Glauner \cite{Glauner:1456329}, evaluation of the \gls{drank} project performed by Olivier Canévet \cite{Canevet:1479357} and Citation graph-based ranking \cite{marian2010}.  

The CDS team also conducts user surveys, whereas only one is publicly available \cite{lemeur}. Another project capturing and analyzing user behavior \cite{Gvianishvili:1295600} is interesting and relevant to this work, as well as the work on Personal Collections by Blacn \cite{yoan}, exploering how personalization affects the user experience.

\subsubsection{d-Rank}
The d-rank project is a vast contribution, exploring how scores from different methods may aggregate into one score used for ranking search results. The \gls{drank} project developed a prototype for CDS and performed several experiments using both generated and real data. The results showed that the rank aggregation has the potential to bring significant improvements in ranking performance to existing item retrieval models for specialized search as compared to the performance of single ranking criterion \cite{vesely2012selection}. The experiments showed this by generating items artificially, assuming that relevant items were more likely to occur on top of the ranked lists. 

Within these experiments, they observed significantly better performance of the aggregated ranking criteria as compared to the performance of the single ranking criteria. Notably, the download frequency and freshness aggregated through linear combination outperformed all individual ranking criteria involved in the experiment, including the download and view frequency, word similarity and freshness \cite{vesely2012selection}.

However, notwithstanding the promising results from their experiments, \gls{drank} was never implemented into CDS. The reason for this is the complexity of configuring \gls{drank} properly, requiring expertise about implementation details, which the team lost after the author completed his Ph.D and left CERN. \\

\subsubsection{Word Similarity Search}
\label{related:word_similarity}
The goal of the work on Solr by Glauner \cite{Glauner:1456329} was to enhance Invenio by bridging it with modern external information retrieval systems. The goal was two folded, the first sub goal was to improve the performance of the search engine, because experiments showed that the internal search engine in Invenio did not scale above 1 million items. 

\begin{figure}
  \centering
    \includegraphics[width=0.8\textwidth]{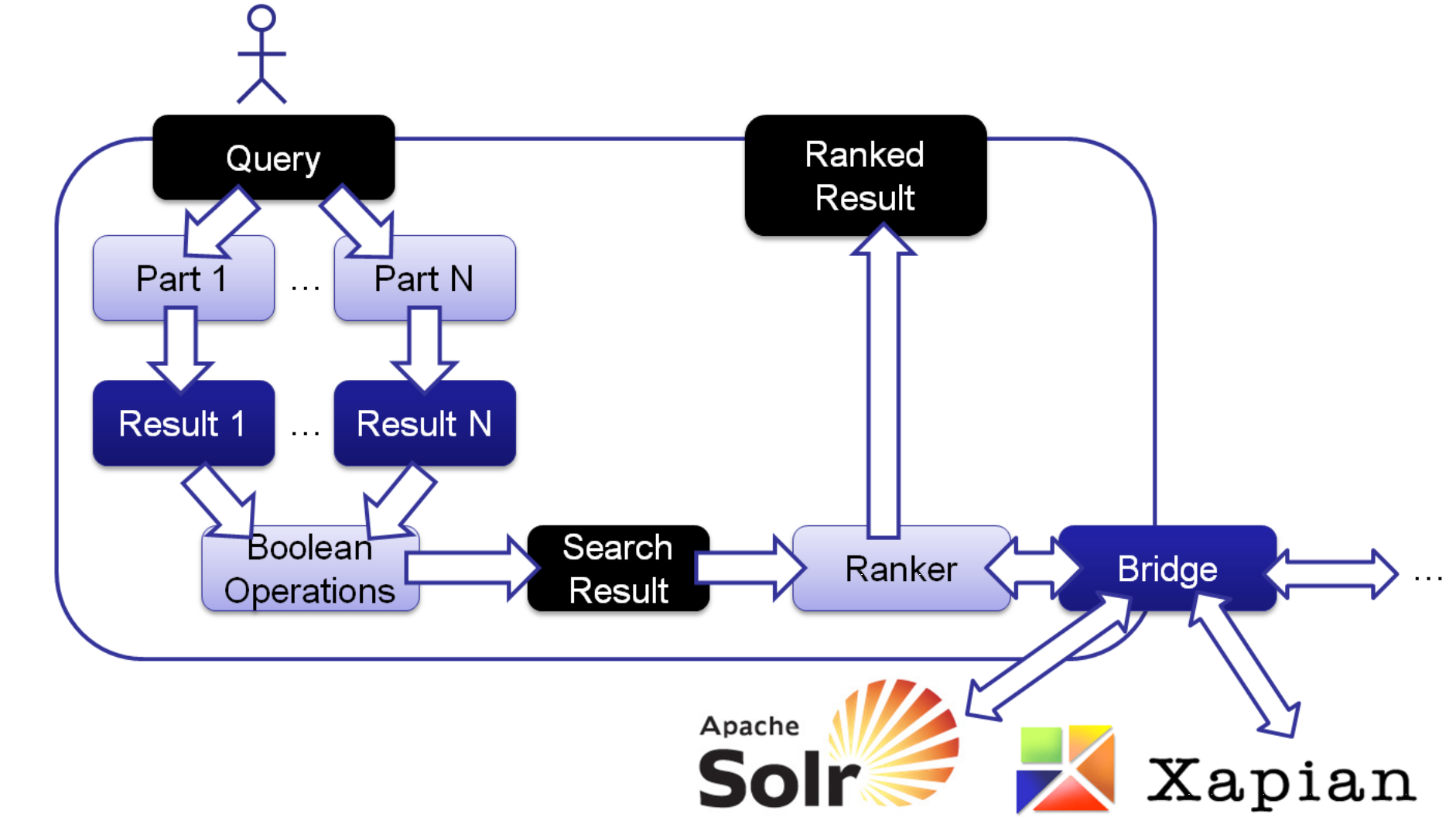}
  \caption{External search and word similarity ranking through the bridge}
  \label{fig:glauner_solrxapian}
\end{figure}

The project compared the search engines Solr, Xapian and ElasticSearch. The experiments showed that Solr outperformed Xapian and that ElasticSearch was too immature for production usage at that time. The bridge is illustrated in figure \ref{fig:glauner_solrxapian}, showing the highly modular architecture regarding search and ranking. 

Invenio performs one independent search per field, each search returns a vector representing the items. Subsequently a Boolean operation (see section \ref{subsec:bgBooleanmodel}) depending on the query is performed on the vectors. At this stage, the result is not ranked. In the next step, the result set is ranked by passing the query and the result ids to the ranker \cite{Glauner:1456329}. 

The bridge enabling word similarity search using Solr is implemented in Invenio today, but not enabled on CDS by default. The reason for this is the complex collections of items in CDS. There exist collections for a broad range of types, such as videos, photos, articles and research results, making a generic configuration difficult to create. Also, within any collection, the items may be very similar. For instance, research on the Higgs Boson tends to contain the same words, the main difference between the items is the number of usages per term, which is not necessarily a good indicator of relevance \cite{vesely2012selection}. \\

\subsubsection{Time Dependent Citation Graph-Based Ranking} 
In a digital library, where the corpus to a large degree consists of scientific papers and articles, one possible for ranking is to use citation based graph ranking. However, research has shown that citation counts alone are not sufficient for this approach. This is due to its inability to account for importance and creation date of the citation targets \cite{marian2010}. The work of Marian \cite{marian2010} presents four types of ranking models based on the citation graph that complements this approach. 

One of the models presents a time-decaying weighting function, where the citation weight of a item is calculated as $weight_i = \sum_{j,j\rightarrow i} e^{-w(t_{present} - t_j)}$, where $t_{present}$ is the present time, $t_j$ is the publication date for item $j$, and ($w \in (0, 1]$) is a parameter that decides the aggressiveness of the ranking model.

A link-based model uses Google's PageRank algorithm to analyze and rank the nodes of the citation graph tree, based on connectivity to the rest of the tree, while a third model combines these two models.

It is also worth noting that the higher completeness and connectivity the citation graph has, the higher value can be placed on these methods. An analysis done on a subset of the CDS corpus by Marian \cite{marian2010} shows that approximately 20\% of the items are not cited by any other item, and 35\% have no references. In addition, on average 28 of 37 references are missing from the CDS per paper (i.e. the referenced items do not exist in the CDS). To overcome this, an external citation weighting algorithm was developed to account for the missing references. A comparison of the link-based ranking with and without the external citation algorithm suggests that the algorithm helps correct the shortcomings of the link-based ranking.

It can be concluded that citation-based graph ranking is a valuable tool, and a significant improvement over the traditional citation count model. Especially the time-dependent link-based algorithm presents promising results, although a few challenges still remain in regard to data correctness of the citation graph (see page 4 of \cite{marian2010}).

\section{Analysis of needs}
This section presents the approach to find the user's needs. The first section presents the relevant findings in a user satisfaction survey from 2011. The following section presents results from the interviews with users of CDS from 2014. The third section presents the interviews with the librarians at CERN, followed by a summary of the findings in these studies. Finally a prototype is proposed, based on the findings collected during the analysis of needs.

\subsection{User Satisfaction Survey}
\label{subsec:search_on_cds_user_survey}

The CDS team ran a User Satisfaction Survey in 2011, covering topics such as the search experience and the submission process, as well as the collaborative tools in CDS \cite{lemeur}. For this project, only the part relevant to the search engine is examined. The survey received 150 answers, out of which 2/3 came from users working at CERN; the remaining users were external users collaborating with CERN.     

25.5\% of the respondents perform daily searches in the CDS service (27\% weekly and 27\% monthly), whereas 20.5\% of the respondents never search. The survey shows that respondents are mostly regular users \cite{lemeur}, it also shows that about 20.5\% of the respondents access the items directly\cite{lemeur}. Direct access to an item is obtained through search engines such as Google or when users send URLs the each other through emails or similar techniques.   

32\% of the users answer that they are always able to find the items they are looking for using the search engine, whereas 52\% of the users do not agree with such a statement. The remaining 16\% are not specified in the result report. The survey also allows the users to write text comments, the most relevant comments from the survey were:

\begin{itemize}
\item The search is often considered as difficult with a user pointing out the need for fuzzy search.
\item Search is considered difficult because ”meta-data quality is poor”.
\item A few users consider that for publicly accessible content, googling site:cds.cern.ch is more efficient.
\item Full-text search is attractive if it can be run in all collaboration collections
in one go”
\end{itemize}

About the ranking preference, 59\% of the users asked to be able to order results according to quality criterias such as downloads, citations, and more. The other 40\% is satisfied with the default ordering of today: \textit{latest first}. 

31\% of users navigate to a specific collection before running a search. 46\% consider it to be useless. Considering the ”Google effect” design - a simple box search and no navigation at all - this result is relatively well balanced. Going back to the simple box might not satisfy about 1/3rd of CDS users. 

43\% of users consider the advanced interface difficult to use. 41\% do not use it at all, and 15\% find it easy to use. That makes 84\% of users who either do not use it or do not like it. This is obviously one area of improvement that must be considered in the next version of CDS. From the logs of user activity, it appears that in 2011, the simple search was used almost 6 times more often than the advanced search (daily average of 19’053 versus 3’707 per day).

A summary of the user study showed that the general impression of CDS is that is feels like a mandatory tool for preserving publications rather than a tool for discovering new content. The search feature is pointed out as one of the biggest sources of frustration, resulting in users moving to services like Google to perform search in CDS with queries like: site:cds.cern.ch:user-query. The users observed that using Google is usually faster and more reliable than the search provided by CDS. One of the users could also report that she always asked for a direct link to items in CDS, searching for it maybe unreliable or impossible. 

\subsection{User studies}
The user studies performed by Blanc \cite{yoan} in 2014 were mainly conducted at the office of the subject or as close to their work environment as possible. The interviews also covered a number of topics not relevant to this work, such as the submission process and the use of baskets, which are explained in section \ref{bg:invenio}. The user study mainly consists of one-to-one interviews with volunteers. The main interest of information to obtain was the usage and perception of CDS, and whether they had any ideas on how changes to CDS could make their work easier. 

The goal of the user study performed by Blanc was to meet with various users of CDS, looking for the following information:

\begin{enumerate}
    \item What is the usage made of Invenio and of any tools related to it.
    \item What is the perception of Invenio.
    \item What ideas do they have that could make their work easier.
\end{enumerate} 

\noindent{
A summary of the feedback from the users of CDS is that CDS feels like a mandatory tool and not the most pleasant one. None of the users described CDS as a service to discover new content. The interviews also discovered issues directly related to search. Some relevant quotes from the interviews conducted by \cite{yoan}:}

\begin{enumerate}
    \item \textit{CDS is clunky}
    \item \textit{Search is slow and unreliable}
    \item \textit{CDS makes me sad, not happy, frustrated}
    \item \textit{CDS is a black hole, hold-all” (in French “fourre-tout”)}
    \item \textit{You have to adapt to CDS}
    \item \textit{Searching CDS is difficult}
\end{enumerate}

Search is the most used feature of CDS and is the biggest source of frustrations \cite{yoan}. For many users, the Google search engine is the usual entry-point to CDS. They learned that with some keywords, it is usually faster and more reliable than the search provided by CDS. Subject M (B.3.13) could tell that she always asks for the exact link of a record in CDS. Searching for it maybe unfruitful or unreliably bringing a duplicate or a different version.

\subsection{Interviewing Librarians at CERN}
\label{subsec:interviewing_librarians_at_cern}
The librarians at CERN have been a part of this project since the beginning; they have provided useful information on users' satisfaction and how the users experience searching CDS. When this project started out, the librarians showed immediate interest in improving the search engine. Meetings with the head of the library, as well as curators and librarians, established the direction and goals for the project. 

The main issue the librarians addressed was how to find items they know are popular but are ranked low due to their publication date or unambiguous words. Older items are hard to find when the search result is sorted by latest first, forcing the user to scroll through several pages to find items published only a few years ago. The challenge is to satisfy researchers who want to find the latest publication in some field, and at time same time satisfy the users who are looking for something particular or maybe something old. The use case of finding the most recent publication should not affect the search engine as it does today. Instead, the search engine should work as any other search engine the users are used to, such as Google or Yahoo, leaving the latest first ranking optional.

The librarians also provided a list of known popular items (such as the \textit{LHC bible} \cite{Breskin:1244506}) that users have trouble finding. An analysis of the items showed that many of the items were old, published 5, 10 or 15 years ago. When users search for these items, they may have to scroll and click on \textit{next page} for 5-10 pages before finding the desired item. A quick simulation of typical queries provided by the librarians used by real users to find these items showed that the desired item could be ranked (\textit{latest first}) as item number 850. Which means that the user has to click \textit{next page} about 100 times in order to find the item, or re-do the search with a more specific query, maybe using Boolean operators to filter out items that does not contain both words for instance. 

\subsection{Lessons learned}
\label{lessons_learned}

Information retrieval is a challenging task, especially for digital repositories containing millions of items. One may identify two tasks for IR systems: 

\begin{itemize}
\item The first is to find information of interest out of a large collection of items.

\item The second is to display the results in a reasonable amount of time in a satisfactory order so that the user can easily find the desired information. 
\end{itemize}

Currently, in addition to sorting methods based on date, author, title, and other meta data fields, several ranking methods are implemented within Invenio such as word similarity (towards the query) or citations (how many times a item is cited by others), which all result in a specific score for the items. 

However, interviews with both real users and librarians, as well as the user survey from 2011, show us that the users are not satisfied by the ordering of the results in CDS; they actually prefer to use Google to find items on CDS. This leaves CDS as a platform mainly for storing items, rather that the goal which is to be a platform for searching and exploring new items from CERN. The analysis are based on the user survey from 2011, interviews with both librarians and real users in 2014 and musings with the core developers of Invenio. The user survey is three years old and should therefore be considered carefully, while the recent interviews tells us more about how the users experiences the present version of CDS. 

CDS is a platform used to serve all publications from CERN combined with harvested content from collaborating organizations. The content of CDS therefore contains a broad range of subjects from articles about findings in physics to guides on subsistence rates for associated members of the personnel, the content are therefore divided into several collections: Articles \& Preprints, Books \& Proceedings, Presentations \& Talks, Periodicals \& Progress Reports and Multimedia \& Outreach. 

One can say that the collections are heterogeneous while the items inside each collection are homogeneous, because most articles abut for example the higgs particle contains the same set of words, which makes the word similarity search less useful. 

The issue of published date could have been solved with a word similarity search, but as mentioned in section \ref{related:word_similarity} on word similarity, the collections of items on CDS makes word similarity searches hard, and sometimes useless.

The most challenging task is, therefore, to make the general \textit{search in all collections} useful. The default ranking in CDS today is \textit{latest first}, which means that if a new publication is added, and it matches a user's search query, the publication will be on top of the search result. The only action taken to solve this issue by default is to show the results grouped by collections, which means that for every search the results are shown in groups by ten (default). For instance if a user search for \textit{Higgs}, the user may get hits from both Articles and Thesis, which means that 20 results are shown, top ten for each collection. It would be interesting to take into account the access frequency of a item in the ranking. 

\section{Proposed Prototype for CDS}
\label{obelix_prototype}

This prototype is based on the aforementioned lessons learned, and one observation from the d-rank project in particular. The d-rank project discovered that the best search results produced were based on a combination of \textit{latest first} and \textit{download history} (popularity) \cite{vesely2012selection}.

An interesting feedback from the librarians at CERN was that users often search for items that their colleagues recommend or that are popular in their field of study. This lead to the idea of introducing collaborative filtering, forming \textit{communities} (clusters of similar users). The idea of forming communities is based on the assumption that similar users have a common interest in items (e.g: users working at \textit{LHC} are more interested in \textit{LHC} items than the users working in IT, whereas users working in IT are more interested in items describing some computer science topic). This assumption is not always true, and for that reason it is important to be able to disable the personalized search easily when searching.

The concept of communities already exist at CERN. The users of CDS are all a part of the \textit{CERN community} and the users of \textit{INSPIRE HEP} are all a part of the \textit{High Energy Particle Physics Community}. However, if the communities are defined by group memberships, it limits the use cases for users who are interested in items produced by another group. By using the existing user groups, the search will not be \textit{personalized} but rather \textit{groupalized}. 

By building communities based on user interactions and relating two users who have used the same item, it is possible to achieve a local optimum for popular downloads (item x is the most popular item in the community of user y). A side effect of this is that users are not put into a community based on where they work, but how they use CDS. Meaning that their recommendations are not necessarily based on where you work, it is based on what you search for on CDS, making the search experience truly \textit{personalized}.

\begin{figure}
  \caption{CDS Trust Model}
  \centering
    \includegraphics[width=0.9\textwidth]{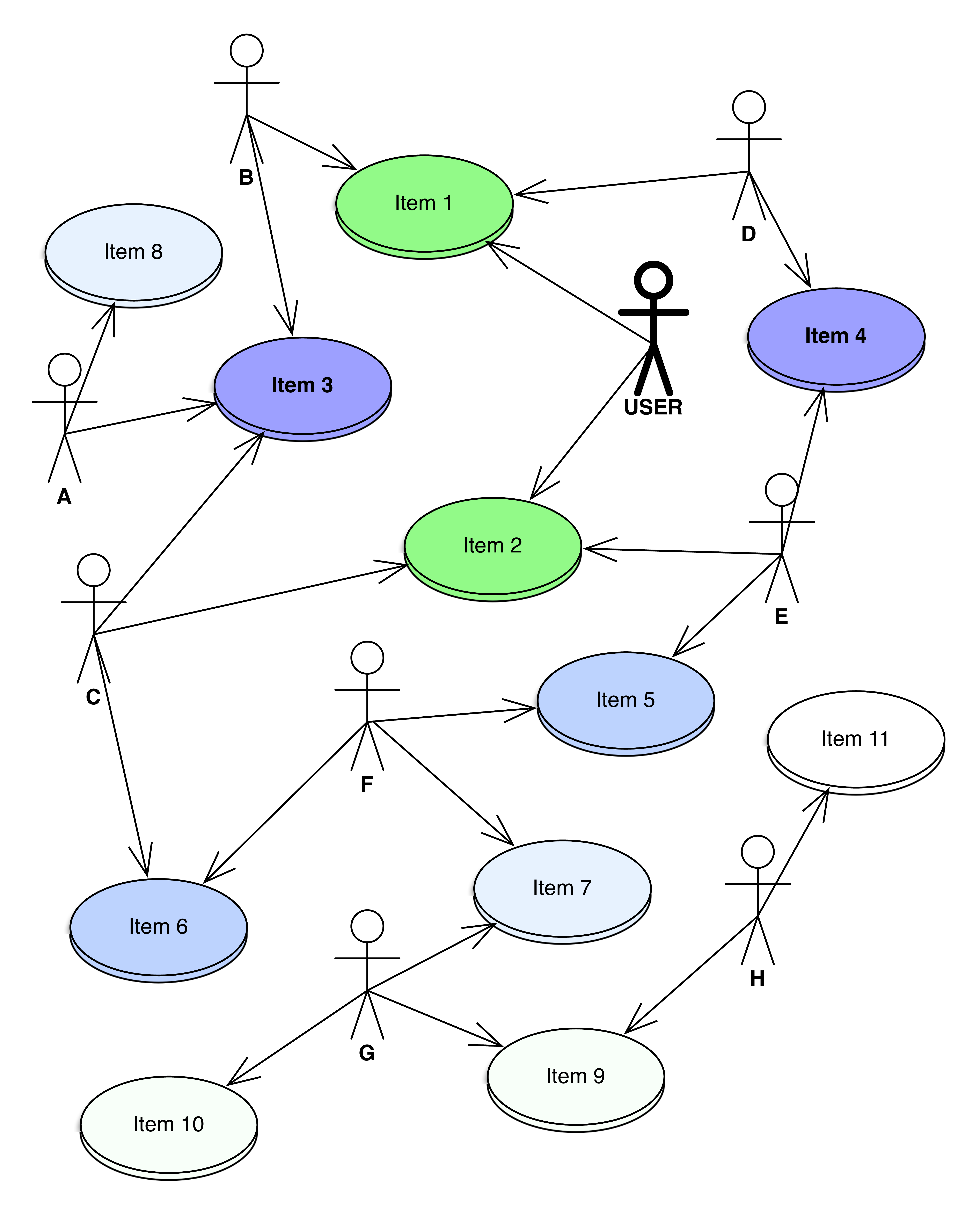}
   \label{fig:obelix_cds_trust_model}
\end{figure}

The relationships between users and items may be illustrated as in figure \ref{fig:obelix_cds_trust_model}. The colors indicate how \textit{far away} the item is from the USER. \textit{Item 1} and \textit{Item 2} are used by the USER himself, marked with green. While \textit{Item 3} and \textit{Item 4} are used by users who have also used \textit{Item 1} and \textit{Item 2}, indicated by a strong blue color for \textit{Item 3} and \textit{Item 4}. In comparison, \textit{Item 5} and \textit{Item 6} are marked with a slightly less strong blue color, indicating that \textit{Item 3} and \textit{Item 4} are recommended stronger than \textit{Item 5} and \textit{Item 6}. When developing the the prototype, a temporary formula was developed to illustrate the scoring:

\begin{equation}
\label{equation_obelix_prototype_calc}
score = \frac{e^n}{{\sum user_d} +1}
\end{equation}

Where $n$ is the number users who have used the item, and $user_d$ is the sum of the distance from the USER to all the users who have used the item. For example, user E has a distance of 2, while user F has a distance of 4. The effect of this difference is that user E has more influence than user F. What follows is that \textit{Item 4} will have a higher score than \textit{Item 5}, because the two users who have used \textit{Item 4} are closer to USER than the users who have used \textit{Item 5}. 

As seen in the table \ref{table:prototype_score_calculation}, the score formula as seen in formula \ref{equation_obelix_prototype_calc} illustrates the weighting of items as described. The formula works for all items in the example, except \textit{Item 9}, which should receive a higher score than \textit{Item 10}. However, the final formula and parameters defining the score based on depth and the weight of each usage is evaluated in the experiment of prediction accuracy in section \ref{predicting_accuracy}.

\begin{table}[h]
\label{table:prototype_score_calculation}
\centering
\begin{tabular}{|l|l|l|}
\hline
\textbf{Item}        & \textbf{Formula} & \textbf{Score} \\ \hline
Item 1        & $ \frac{e^3}{((0+2+2) + 1)}$ & 4.01 \\ \hline
Item 2        & $ \frac{e^3}{((0+2+2) + 1)}$ & 4.01 \\ \hline
Item 3        & $ \frac{e^3}{((2+2+3) + 1)}$ & 2.51 \\ \hline
Item 4        & $ \frac{e^2}{((2+2) + 1)}$ & 1.47 \\ \hline
Item 5        & $ \frac{e^2}{((4+2) + 1)}$ & 1.04 \\ \hline
Item 6        & $ \frac{e^2}{((4+2) + 1)}$ & 1.04 \\ \hline
Item 7        & $ \frac{e^2}{((4+6) + 1)}$ & 0.67 \\ \hline
Item 8        & $ \frac{e^1}{((4) + 1)}$ & 0.54 \\ \hline
Item 9        & $ \frac{e^2}{(6+8) + 1)}$ & 0.18 \\ \hline
Item 10       & $ \frac{e^1}{((6) + 1)}$ & 0.38 \\ \hline
Item 11       & $ \frac{e^1}{((9) + 1)}$ & 0.27 \\ \hline
\end{tabular}
\caption{The scores for the items in figure \ref{fig:obelix_cds_trust_model} calculated.}
\end{table}

\chapter{Obelix Design}
This chapter includes the design process and architecture of Obelix, a new recommendation system built to add personalization to the search engine. Obelix is generic and not only works with CDS, but is also compliant with most digital libraries and other IR systems.

\section{Introduction}
The product of this project is a personalized search engine based on the new recommendation system Obelix. The design for Obelix is based on the proposed prototype in section \ref{obelix_prototype}. 

Obelix is a complement to an existing search engine like Solr or ElasticSearch, which focuses on search using keywords. Obelix recommends items based on users relationships with items and other users as mentioned in section \ref{obelix_prototype}. The recommendations from Obelix are used to boost items in the \textit{Original List}, attempting to improve the search experience by re-ranking the results.

Obelix is a standalone software (see section \ref{sec:integratedvsstandalone}), compatible with any IR system through its open APIs. By making Obelix standalone and available through a REST API, the integrations with other IR systems becomes independent of programming language. All actions are available through the REST API. However, in order to obtain the best possible performance, it is recommended to make use of a key/value store and a queue as the interface for Obelix to the integrated IR system, as described in section \ref{sec:high_performance_setup}.  

\section{Requirements}
\label{sec:obelix_requirements}
When developing a system that integrates with a service like CDS with more than 40 000 unique active users \cite{Gvianishvili:1295600} and more than 1.4 million items, stability and performance are important factors for implementing a successful design. For instance, it is not acceptable if Obelix becomes a bottleneck in the search process or if a bug in Obelix leads to downtime of the service. 

The requirements for Obelix are split into two lists, the first list containing requirements from the CDS team in order to run Obelix in production. The second list contains requirements made with the goal of making Obelix generic and compatible with any IR system.  

\subsection{List of requirements from the CDS team}

\begin{description}

\item[Requirement 1] \hfill \\ The personalized search feature should be easy to enable/disable. 

\item[Requirement 2] \hfill \\ The users should be able to expect the same speed performance as today, the service should not become noticeable slower.  

\item[Requirement 3] \hfill \\ The integration with Obelix should handle errors in Obelix and make sure CDS is functional even if Obelix is down.

\item[Requirement 4] \hfill \\ Obelix needs to be easy to configure and install to make it possible for the IT staff to maintain the installation on their own. 

\item[Requirement 5] \hfill \\ Obelix needs to run on the servers provided by CERN IT, running Scientific Linux.

\item[Requirement 6] \hfill \\ Obelix needs to support importing millions of interactions before starting (10 years of data) to avoid cold start.

\item[Requirement 7] \hfill \\ Obelix has to be easy to evaluate, in order to decide if it should be installed permanently.

\item[Requirement 8] \hfill \\ Obelix has to be able to visualize the graph built to generate recommendations. This requirement is essential in order to explain the underlying data structure and how the system works for other developers, stakeholders as well as IT operations.

\item[Requirement 9] \hfill \\ Obelix needs to be self-learning and improve over time. in order to be able to start without information.

\end{description}
\hfill

\subsection{List of requirements to make Obelix generic}

\begin{description}

\item[Requirement 10] \hfill \\ Obelix has to be generic in the way it handles item usage. It should support not only \textit{downloads} or \textit{views}, but also any other interaction such as \textit{shares} or \textit{likes}. 

\item[Requirement 11] \hfill \\ Obelix has to provide logs.
 
\item[Requirement 12] \hfill \\ Obelix has to ensure uniqueness for users, relationships, and items. 

\item[Requirement 13] \hfill \\ Obelix needs to provide an open API for the IR system to integaret with.

\item[Requirement 14] \hfill \\ Obelix has to communicate with IR systems efficiently.

\end{description}

\section{Integrated vs Standalone}
\label{sec:integratedvsstandalone}
Invenio is built using Python so in the scope of this project; it would be enough to support a Python integration. However, this project aims at supporting more than only \gls{cds}; Obelix should be a generic and useful recommendation system to add personalization to any search engine. 

Nevertheless, there are some challenges of decoupling Obelix from the IR systems. The know-how of the integration and configuration of Obelix needs to stay within the team of developers of the IR system, in order to manage upgrades and to optimize the parameters as the IR system grows. Another issue related to the decoupling is complexity and required resources for the installation and integration, e.g. if Obelix were developed as a module inside of Invenio, it would work out of the box when installing Invenio.   

However, by decoupling Obelix from the IR system, it is possible to continue the development of Obelix without interrupting the development of the IR system and vice versa, as long as Obelix maintain a stable API. 

Another important aspect is the importance of controlling the environment where Obelix is running, both in the case of daily maintenance and to ensure Obelix has the required resources to perform adequately. By running Obelix standalone, an abstraction layer is introduced in the interface between Obelix and the IR system. The abstraction allows changes to internal methods of Obelix without the need for amending the code of the IR system. It also makes dependency management simpler, as long as Obelix is defining the environment; there will be no dependency issues like incompatible versions of underlying packages between the IR system and Obelix. 

By making Obelix standalone, it is simple to integrate Obelix with any other IR systems. This will allow more projects to adopt and make use of Obelix for personalizing search or other creative use cases. There are only two things the IR system will need to integrate, the usage of items (view, like or some other interaction desirable) and the re-rank as part of the current search engine. The choice of making Obelix standalone and decoupled from the IR system is the best option in order to match the requirements set by both the CDS team and the requirements to make Obelix available for other IR systems.    

\section{Collecting data}
Obelix generates recommendations based on the interactions between users and items within a given IR system. There are three events of interest: 

\begin{itemize}
\item Item usage (\textit{view} or \textit{download} in the case of CDS).
\item \textit{Click position} in search result (for use in statistics).
\item Relating search result to item usage (for use in statistics).
\end{itemize}

It is necessary to capture these events as the user interact with the system in order to continuously improve the results of the algorithm and thereby provide better search results iteratively. 

However, it is only required to capture the item usage for Obelix to function. The \textit{click position} and the relationship between a search result and an item usage are only relevant for evaluation. Evaluation in this case refers not only to the experiments conducted in chapter five, but also as a tool tune Obelix over time. For that reason, it is recommended to capture all three events in order to collect useful information for use in statistics and tuning the performance.

\begin{figure}
  \caption{Simple data flow}
  \centering
    \includegraphics[width=0.8\textwidth]{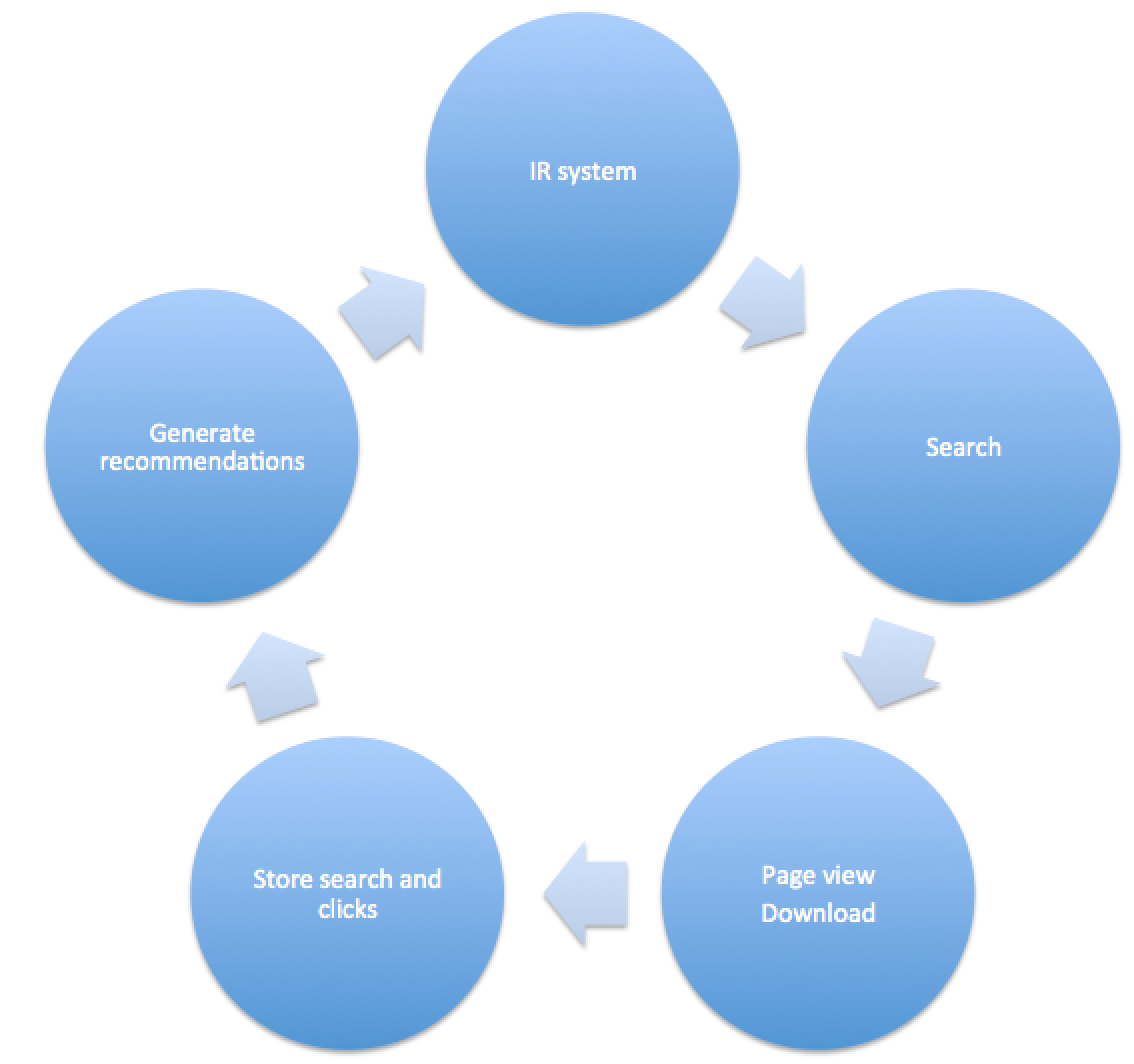}
   \label{fig:obelix_simplest_data_flow}
\end{figure}

In order for Obelix to work, data about the users interaction (views, downloads, and searches) is required, which is provided by the given IR system. Obelix and the IR system needs a two way communication in order to generate the recommendations and provide the recommendations to the search engine. An illustration of the simplest data flow may look like figure \ref{fig:obelix_simplest_data_flow}, the data flow is customizable and defined by the integration with the specific IR system. The data flow typically starts when a user search using the search engine and then click on an item. The \textit{click position } and search result are then stored in Obelix and used to generate recommendations for that user. The next time the user search using CDS, the previous search and click affects the new search result by re-ranking according to the generated recommendations.

\section{Recommendation Algorithm}
\begin{figure}
  \centering
    \includegraphics[width=1\textwidth]{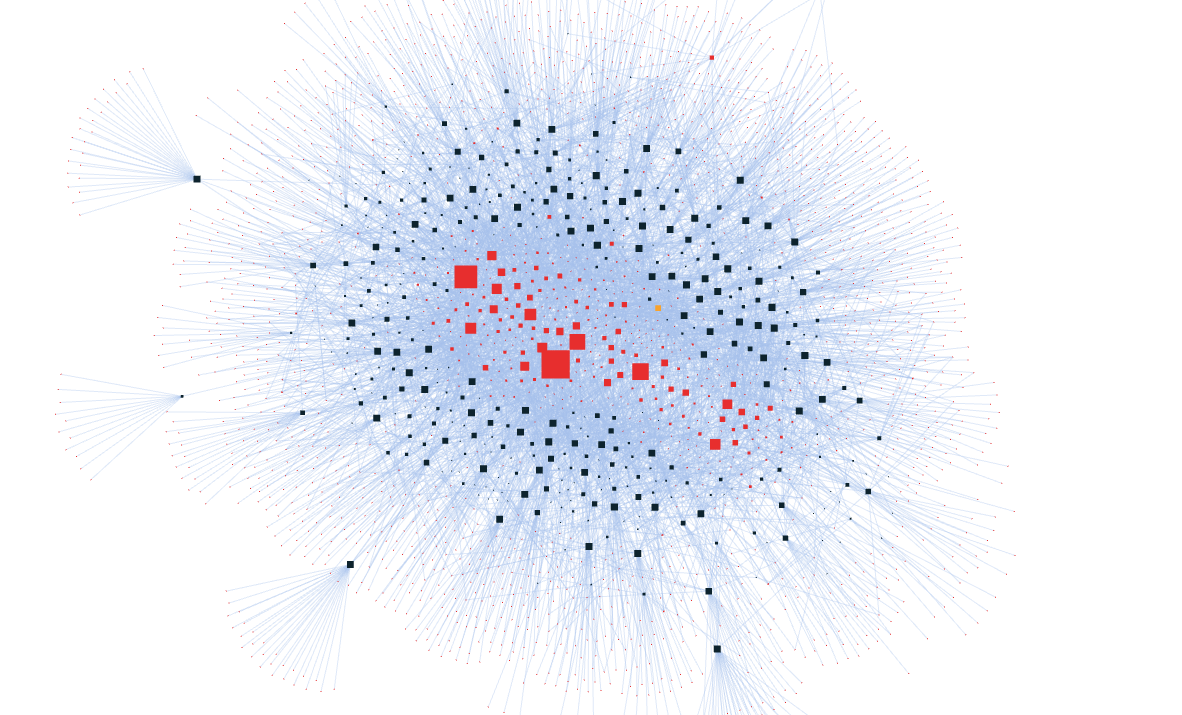}
  \caption{Small set of users and items related}
  \label{fig:small_set_of_users_graph}
\end{figure}

The calculations are based on the concept of collaborative filtering (see section \ref{sec:background:collaborative_filtering}). Every user is represented as a single node in the graph, width edges to items they have used, visualized in figure \ref{fig:small_set_of_users_graph}, where red nodes are items, black nodes are other users and the yellow node is the current user. The visualization graph is built using collected data from CDS in production. 

The core idea is to traverse the graph from a starting node, representing the user (the yellow node) to a given depth. The black nodes are other users, while the red nodes are items. The size of the red nodes is set by their score, large nodes have large a high score. The score is calculated based on how many users have read the items found during the traversal, combined with how far the item is from the current user (distance measured by number of edges traversed). The closer the item is to you (shortest distance) the better score it will get. The final score for a item is described by this formula: 

\begin{equation}
score = \frac{e^n}{{\sum user_d} +1}
\end{equation}

Where $n$ is the number of users who have used this particular item, and $user_n$ is the distance from the current user to the user who have used it. The distance is defined as the shortest path from the current user. The algorithm traverse the graph using \gls{bfs}, discovering the items and users who are close to the current user first. 

\begin{algorithm}
\caption{Scoring algorithm}
\begin{algorithmic}[1]
\Require A directed $G = (V, E)$ of order $n > 0$. A
  vertex $s$ from which to start the search. The vertices are numbered
  from $1$ to  $n = |V|$, i.e.~$V = \{1, 2, \dots, n\}$.
\Ensure A list $D$ of distances of all vertices from $s$. A tree $T$
  rooted at $s$.
\State $Q \gets [s]$\label{alg:BFS:initialize_queue_visit_nodes}\Comment{queue of nodes to visit}
\State $D \gets [\infty, \infty, \dots, \infty]$\Comment{$n$ copies of $\infty$}
\State $D[s] \gets 0$
\State $T \gets [\,]$\label{alg:BFS:initialize_empty_tree}
\While{$\left|Q\right| > 0$\label{alg:BFS:while_loop:non_empty_queue}}
  \State $v \gets dequeue(Q)$
  \For{\textrm each $w \in v$\label{alg:BFS:explore_neighborhood}}
    \If{$D[w] = \infty$\label{alg:BFS:marking_vertex_as_visited}}
      \State $D[w] \gets D[v] + 1$
      \State $enqueue(Q, w)$
      \State $calculateScore(w)$
      \State $append(T, vw)$\label{alg:BFS:while_loop:append_to_tree}
    \EndIf
  \EndFor
\EndWhile
\State \Return $(D, T)$
\end{algorithmic}

\end{algorithm}

\section{Scalability}
In order to meet the requirements on performance, Obelix has to be built based on a flexible and scalable architecture. It is important to minimize and remove any potential bottleneck of the data flow. Flexibility is achieved by splitting Obelix into smaller modules working together, making it possible to create a new or replace an existing module if it does not perform as expected.

It is possible to achieve scalability by scaling horizontally or vertically (see section \ref{background_scaling}). In the case of Obelix, scaling vertically is obtained by making the number of threads and amount of memory  configurable. At any time, Obelix may be restarted with more threads and memory, continuing from the state before the restart. Horizontal scaling is obtained by starting several nodes with Obelix installed on the same subnet (physical or virtual); the Obelix instances running on different nodes will discover each other and distribute the load automatically. However, there is always a single master and zero or more slaves. Compared to other master-slave replication setups, Obelix handles write requests on all nodes, so there are no need to redirect the write requests to the master. A slave handle writes by synchronizing with the master to preserve consistency. All updates will propagate from the master to the slaves eventually, so the write from one slave may not be visible on all slaves immediately. 

\section{Obelix Architecture}
This section explores the architecture of Obelix, including the concepts and technologies used, as well as interfaces available through REST or key/value. 

\subsection{Building blocks}
\label{building_blocks}

\begin{figure}
  \caption{Architecture overview}
  \centering
    \includegraphics[width=0.8\textwidth]{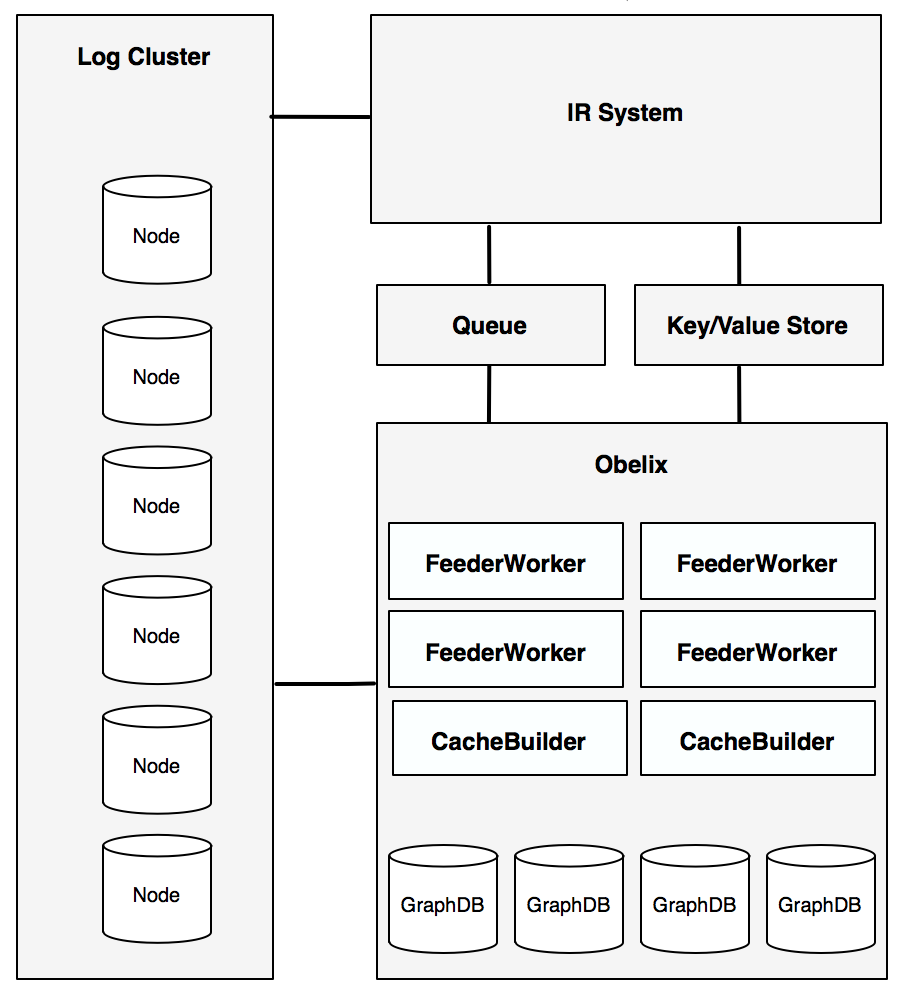}
   \label{fig:obelix_architecture_overview}
\end{figure}

Figure \ref{fig:obelix_architecture_overview} shows an overview of Obelix and the recommended interfaces (queue and key/value store) with an IR system. Each component is described in section \ref{building_blocks}.  

\subsection{GraphDB}
\label{subsubsec:graph_database}
The graph database is the core of Obelix, storing all relationships between users and items. The graph is used for traversal, starting from given user, using Breadth-First, to some depth \textit{x}, which is configurable. The algorithm uses the paths and distances in the graph to calculate the recommendations for a given user. For that reason, Obelix needs to traverse the graph efficiently and maintain a consistent graph without duplicates. In addition, the graph needs to support batch inserts or something equivalent in order to meet requirement 6. The graph also needs to distribute the workload across multiple instances in order to scale. 

There exist several graph databases such as Neo4j, FlockDB, OrientDB, Titan and DEX. Nevertheless, the different graph databases serve different purposes. For instance, FlockDB, which is developed by Twitter, focuses on direct edges instead of traversal because they care only about which users a user follow, not whom a user follows that other users follow.   

When choosing a graph database for Obelix, traversal is this most important feature to consider. The reason for this is that traversal is the key to generate recommendations, and the graph to traverse may be large, as in the case of CDS. Inserts of new nodes and relationships only occur when users interact with the system, making the insert speed less critical. Based on the conclusion of a comparison on Neo4j, OrientDB, Titan and DEX in 2014, performed by Jouili and Vansteenberghe, the choice is simple. 

\begin{quote}
Based on our measure, the database that obtained the best results with traversal workloads is definitely Neo4j: it outperforms all the other candidates, regardless the workload or the parameters used \cite{JouiliV13}.
\end{quote}

Another important aspect of Neo4j is that it is Open Source under GPL license, allowing us to package Neo4j with Obelix. That Neo4j is open source also enables us to contribute if we discover any bugs or potential for improvements.

\subsection{Queue}
Obelix maintains two queues; one for inserts and another for building recommendations. 

\begin{description}
\item[Event queue] \hfill \\
When Obelix receives a new event, such as a view or a download, the event is pushed on to the \textit{event queue}. The number of workers which pop from the \textit{event queue} continuously is configurable in Obelix. A pop from the \textit{event queue} triggers an insert into the graph database. When an insert operation is complete, the \textit{user id} from the event is pushed to the \textit{recbuild queue}.
\item[Recommendation queue] \hfill \\
There are a configurable number of workers RecBuilders which pops the \textit{Recommendation queue}, triggering a traversal starting from the user indicated by the event. This traversal produces a list of recommended items for the user, which is then stored in the key/value store.
\end{description}

The reason for handling these events with queues is to increase the capacity of Obelix to receive such events. There is no reason for the user to wait for a response from the traversal, it will only affect the user next the user performs a search. This architecture assumes that the insert of new events is not time critical since it will insert new events on a best-effort basis. 

Obelix supports a simple native built-it queue to make Obelix simple to configure. However, there are three major drawbacks of using this native queue. The first is performance, the implementation is just a simple ArrayList in Java, without any optimizations. The second is that the queue data is only stored in memory, which leads to data loss if Obelix is restarted or the host is restarted with data in the queue. The third is scalability; the native queue does not support any distribution of data between Obelix hosts. 

The native built-in queue may be enough for small applications or to test Obelix, but for serious applications like CDS, most sophisticated implementations are required. However, the queue methods are defined as an interface in Obelix, allowing the user of Obelix to implement the queue of choice quickly. For CDS, Redis was chosen as the queue engine for two reasons, efficient redis libraries exist for Java, the other reason is that CERN IT supports Redis on their servers by easily.

\subsection{Key/Value Store}
Obelix uses a key/value store for caching and statistics. When the process of generating recommendations for a user finish, the result is stored in the cache (key/value store). Using the cache enables quick lookup when the IR system asks for the recommendations for a given user (constant time). 

As for the queue, Obelix provides a simple native implementation for the key/value store in memory, with the same drawbacks as for the queue. By using the native implementation, there is a risk of data loss when restarting Obelix and no support for scaling across several machines. 

The key/value store is an interface in Obelix, the same way as the queue, allowing the users of Obelix to use the backend of their choice. The major key/value stores on the market are Redis, Voldemort, Dynamo and Riak.

The choice for of Key/Value Store for the CDS implementation is Redis, mostly because of the well documented and fast implementation of the Redis binding. 

\subsection{Obelix REST-API}
\begin{figure}
  \caption{Obelix Data Flow using REST}
  \centering
    \includegraphics[width=0.8\textwidth]{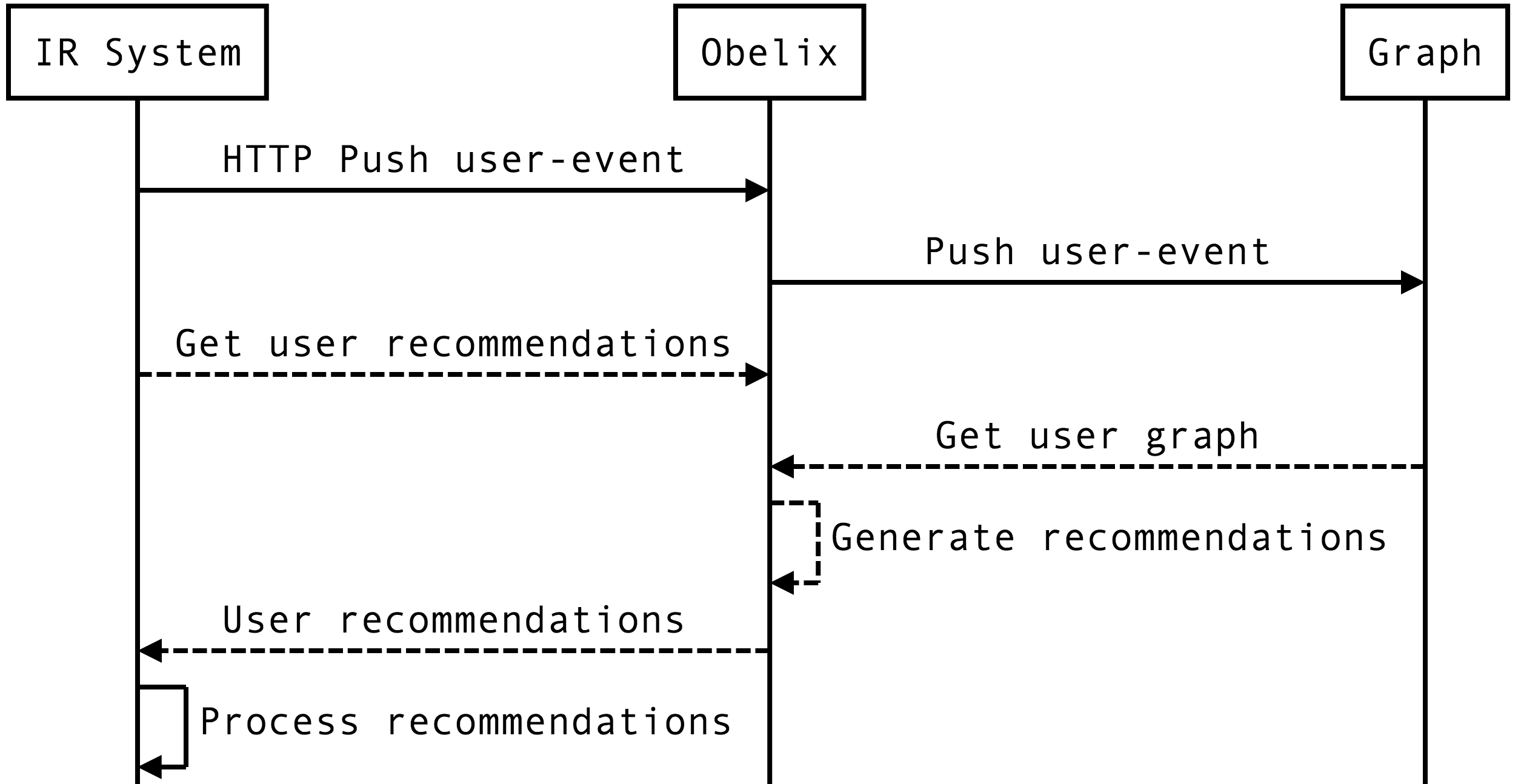}
   \label{fig:original_obelix_arch}
\end{figure}

Obelix offers a complete REST API for all its features. The IR system and Obelix communicates directly over HTTP, as shown in figure \ref{fig:original_obelix_arch}. Every user interaction such as view or download results in an HTTP Request to Obelix. Similarly, when a user in the IR system asks for recommendations, an HTTP request is made, making Obelix generate the recommendations on-the-fly or using the cache (optional).  

This approach worked well at first; it worked well with a million documents and tens of thousands of users. However, when the parameters inside Obelix were tuned to make the recommendations better, a magnitude more recommendations were produced. From a few hundred to hundred thousands of recommendations. The large amount of recommendations introduced two problems. The first problem was that the time required to generate recommendations for one single user increased from a few milliseconds to seconds. The second problem was due to the huge amount of data transfer (megabytes) of JSON objects filled with recommendations sent back and forth, making the systems slow and thereby failing requirements provided by the CDS team.

\begin{figure}
  \caption{Comparison between usage of Redis and API}
  \label{fig:seconds_process_events}
  \centering
    \includegraphics[width=1.0\textwidth]{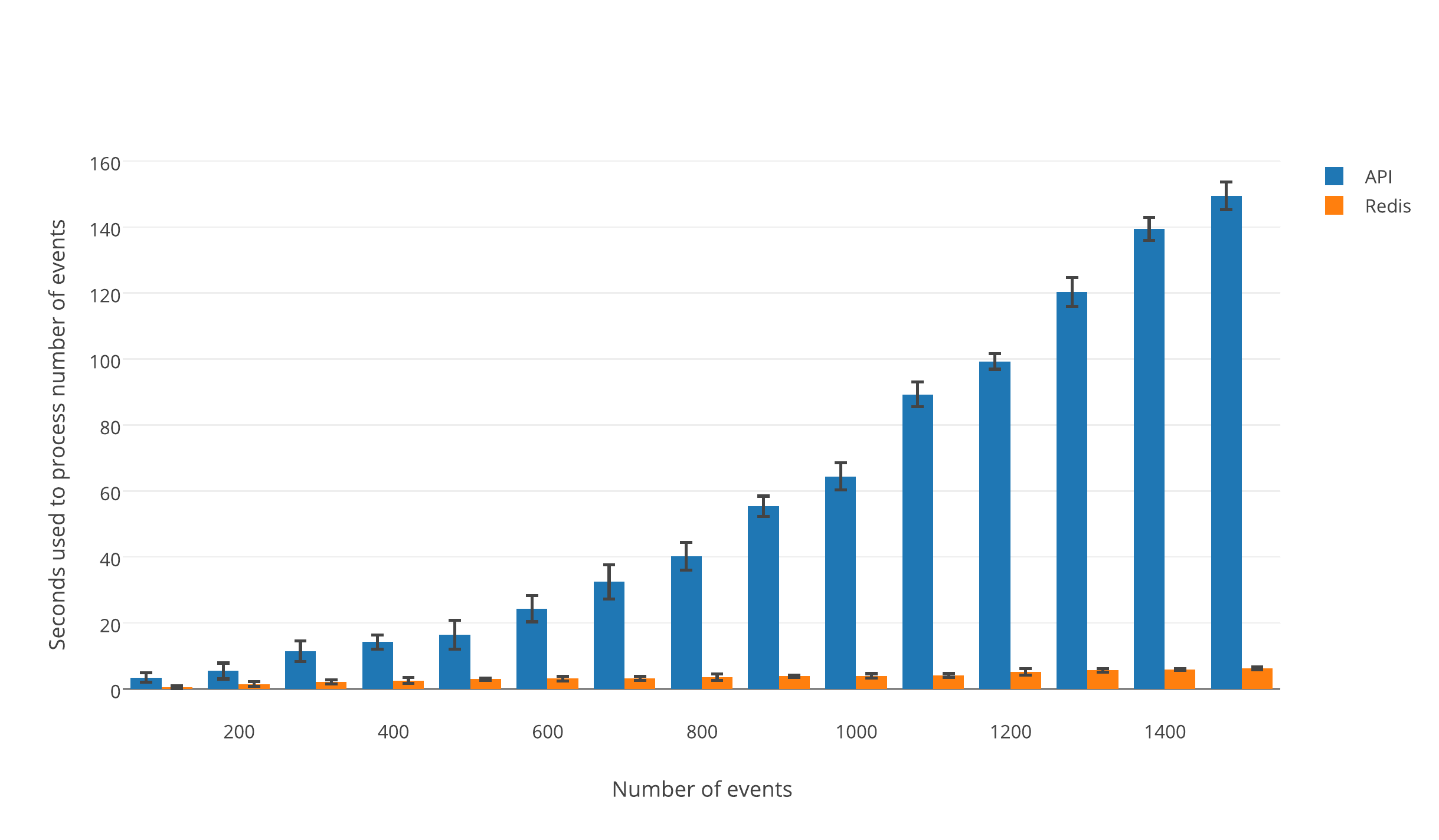}
\end{figure}

Again, as with the native queue and key-value store, using the REST API may be enough for small applications or testing (The REST API was used a lot while debugging Obelix). However, for a service like CDS serving thousands of users every day, it represents a bottleneck. The performance is partially solved by enabling the cache, but the overhead of the HTTP headers and passing data through the HTTP protocol was still noticeable. Especially with HTTPS, forcing a negotiation of the SSL certificate, which takes time. The difference between using the REST API and Redis as in the \textit{High-performance setup} is shown in figure \ref{fig:seconds_process_events}. 

Because of requirement 2, it was necessary to explore more efficient methods for communicating between Obelix and the IR system, leading to the next section on high-performance setup with Obelix. 

\subsection{High-performance setup}
\label{sec:high_performance_setup}
In order to achieve high performance, a caching layer was introduced between the IR system and Obelix. The caching layer made it possible to cache the recommendation generated, as well as make use of a queue for inserting new data into the graph. This architecture allows Obelix to scale, because Obelix handles every new input on a best effort basis, by popping from the queue whenever Obelix is ready to process a new event. The queue makes it possible to handle new events without slowing down the user experience. 

The reason this architecture does not slow down the user experience is because the user never has to wait for any processing by Obelix. The time complexity of both insert of new events and fetching the recommendations are in constant time O(1). Constant time is achieved by only pushing new events on the queue, without having to wait for any reply, except for an answer from Redis. The constant time for fetching recommendations is done by using the \textit{key/value store}, which essentially is a hash table with direct lookup. However, the time complexity may increase when scaling the infrastructure of the system, but that is outside of the scope for this project. 

\begin{figure}[htbp]
  \centering
      \includegraphics[width=0.9\textwidth]{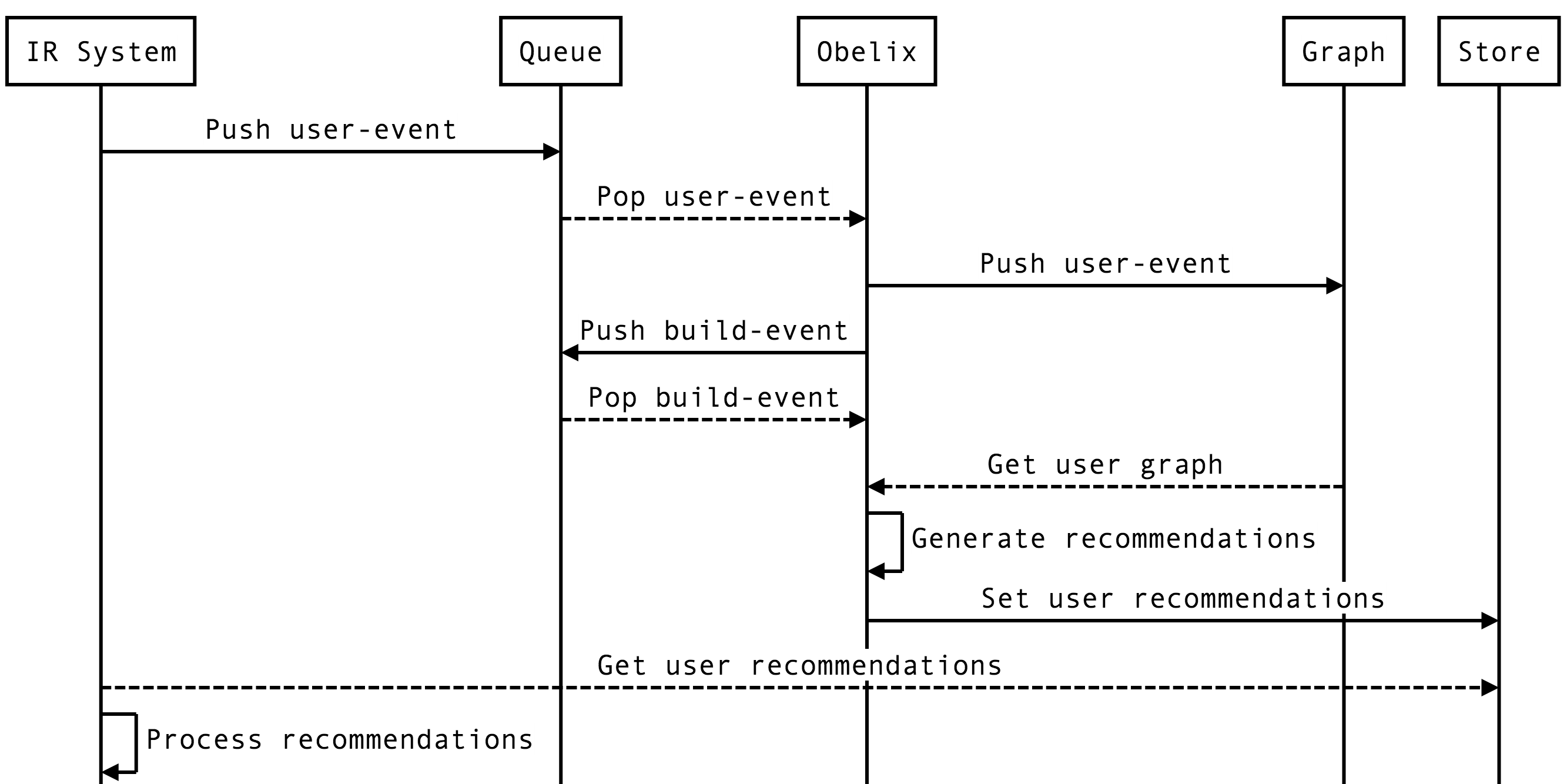}
  \caption{Data flow using queues and a key/value store}
\end{figure}

\subsection{Programming language}
Based on the requirements for this project (see section \ref{sec:obelix_requirements}), the programming language for Obelix needs to be fast and reliable (requirement 2 and 13). It needs to be easy to install and maintain (requirement 4), able to run on Scientific Linux (requirement 5), as well as generic and extendable (requirement 10). 

There are many languages which are matching these requirements, such as C++, Java, Python (with use of C-bindings). However, there are other features worth considering when choosing the programming language:

\begin{itemize}
\item The availability of libraries and data structures.
\item The community of the language in the field (search and databases)
\item The packaging methods available to distribute the program
\end{itemize}

Based on these features, the choice of language for this project is Java. The reason for this is the tremendous number of libraries and bindings available. Even more critical, Neo4j which is the chosen default graph database (see section \ref{subsubsec:graph_database}) supports an embedded version for use within Java applications to obtain the best performance possible with Neo4j. The community for Java and search applications is also a factor (Solr, ElasticSearch, and Neo4j are all written in Java). Another winning point is the packaging of Obelix, which is done very elegantly using Java. Deploying a new version of Obelix only require one jar-file to be deployed.

\subsection{Packaging and hosting}
Package and distribution of Obelix are trivial, the project is built using maven and packed into one single JAR-file. The deployment is as easy as  just copying the file to the server. There is no need for an application server like Glassfish or Tomcat. Obelix comes with a built-in Jetty server. All that the user needs to do is to run the command: \textit{java -jar Obelix.jar}, optionally passing parameters like specifying where Redis is located.  

\subsection{Logging}
In order to evaluate Obelix, it is necessary to log queries and results in order to later create statistics. There exist two categories of statistics needed for Obelix. 

The first is user behavior (usage) and search results. A proper integration (hooks) with the IR system is required to make this possible. The integration needs to report back to Obelix what the users click on (usage) after searching as well as how many records the user had to look through in order to find this particular item. In this work, CDS is the IR system integrated with Obelix. The second category of statistics includes how well the system performs in general, how many recommendations can Obelix produce per minute, how many \gls{qps} can Obelix handle, etc. More information on the statistics is described in chapter \ref{experiments}.

\section{Integration with an IR system}
For Obelix to work, an integration with the IR system is required. The recommendations may be used alone, or combined with another score. In the use case of CDS, the only available ranking to combine with the recommendations is latest first. The approach is to give all items in the original result a score from 0.0 to 1.0, and then join this result with the recommendations. Only items that are in the original search result are included in the final re-ranked result. 

CDS will only communicate with Obelix through Redis, when a user enter CDS and search using a query, CDS returns a set of items (matching keywords) ranked by latest first. Then CDS asks Redis for recommendations for that particular user, loops through the result set and boost those items which are recommended. There is also a API available for the IR System to use, but it is slow compared to the Redis interface, there is no cache and the recommendations are calculated on the fly. However, since the calculations are generated on the fly, it is possible to limit the queries by time, enabling simulation back in time. Using the HTTP API with time ranges, it is possible to query what would the recommendations be at a specific point in time, this is used in the evaluation of Obelix. 

\begin{figure}
  \caption{Re-rank process}
  \centering
    \includegraphics[width=0.8\textwidth]{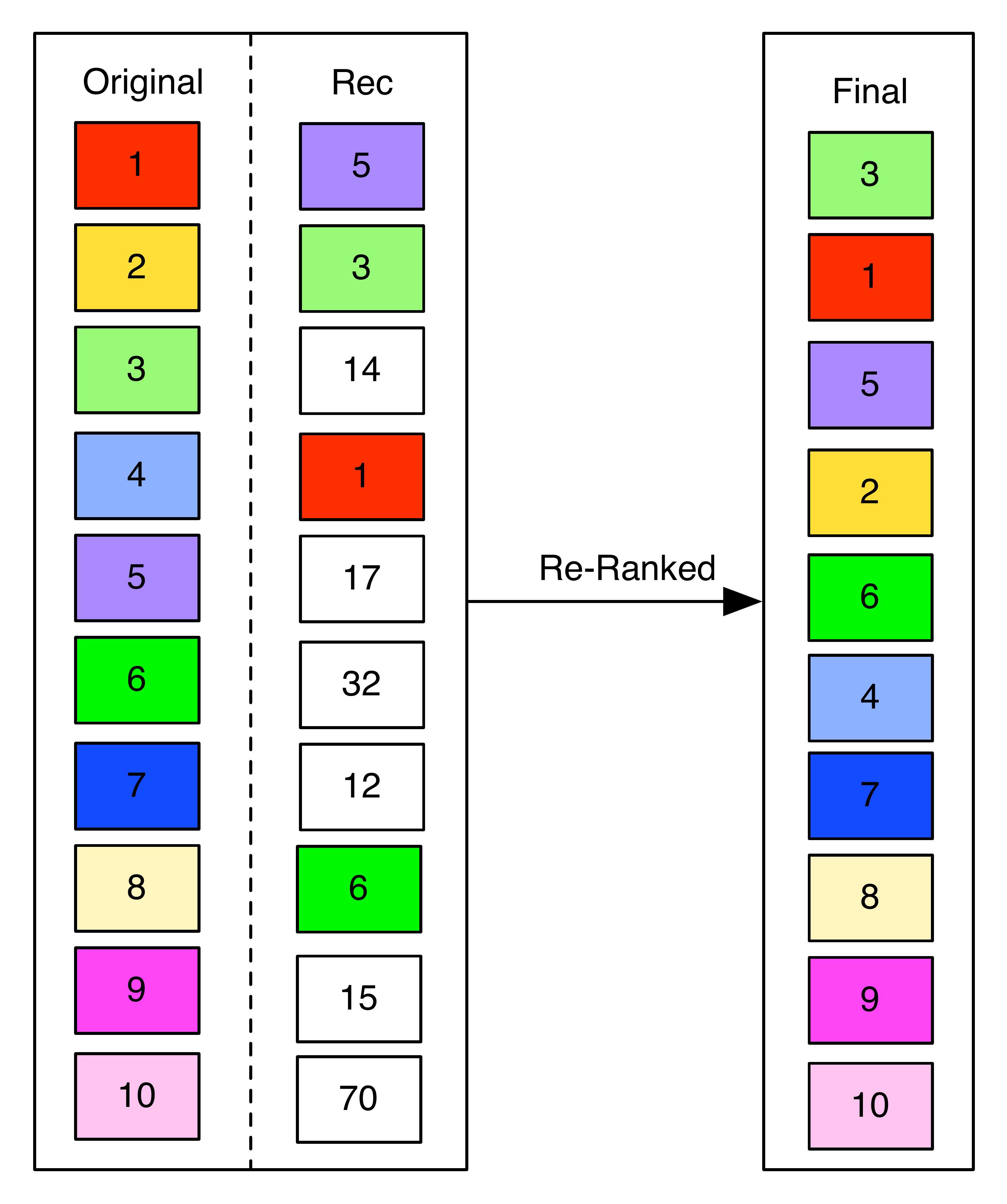}
   \label{fig:re-rank-process}
\end{figure}

An important detail to notice is that in the integration chosen for CDS, Obelix will not introduce or remove any items from the original result, it will only re-rank them, in other words, change the order of the search result. The re-rank is introduced as a second step in the query process, first the IR system queries for all matches, then ranked by a given ranking method, then it is re-ranked by Obelix, taking the original score into account. The process is visualized in figure \ref{fig:re-rank-process}.

\chapter{Experiments}
\label{experiments}

\section{Introduction}
This chapter describes the experiments performed to evaluate and improve Obelix to meet all requirements and perform adequately. Evaluation plays a vital role in information retrieval research, typically in the form of an experiment showing changes in the retrieval performance searches compared to a baseline. In the context of this project, two major areas are valuable to test, the efficiency and effectiveness of Obelix and the personalized search result. 

Although several communities and forums exist for evaluation of information retrieval systems, namely TREC, NTCIR, FIRE and CLEF. However, to the best of our knowledge, no specialised track for the domain of High Energy Physics or scientific publications have been established \footnote{http://trec.nist.gov/tracks.html}. For that reason, it is necessary to create the test sets as a part of this project. 

In order to create the data set needed to evaluate Obelix, existing logs and meta data from CERN were used. The CDS team has stored more than ten years of historical data since the first launch of the service. The team kept the logs from the web server, containing user's interaction, searches and clicks. The new data set developed for this evaluation is built using data from real user's interactions, and this chapter describes the methods and processes used to construct this data set as a resource for evaluation. 

The experiments are split into two categories, offline and online experiments. The offline experiments are performed using logs and existing data, whereas the online experiment are performed in production, with live user interactions and data.

\section{Offline experiments}
\label{sec:evaluation_offline_experiments}

\subsection{Introduction}
Simulating user's behaviour performs an offline experiment that interact with the recommendation system, using pre-collected data sets such as logs. In doing so, one assumption is obligatory; the user's behaviour is similar enough before and after the new system is introduced \cite{microsoftevaluation}. Offline experiments are attractive because they do not require any interaction with real users, making it possible to test a wide range of algorithms and parameters efficiently, without having to wait for feedback from users. 

Although offline experiments are an efficient process for filtering algorithms and parameters, the downside is the narrow set of questions possible to ask. It is for this reason not a sufficient evaluation method for the entire system. As a result, the goal of offline experiments is to filter out inappropriate methods, algorithms and parameters. Offline experiments leave us with a relatively small set of unanswered questions about the performance and accuracy of the system to test with the more costly online experiments. A typical workflow is to select an algorithm and filter out the worst combination of parameters and then continue to the next phase. 

In order to evaluate the algorithms and parameters offline, it is necessary to simulate the process where the system makes recommendations and then try to predict what users click. In this project, historical data (logs) from CDS are used to make such recommendations and predictions. Ten years of logs are available from the web server, consisting of all user interactions with CDS since 2005. An important property of Obelix is how it will affect the search results CDS. Since the main purpose is to reorder the results, not recommend arbitrary items to the user, this also needs to be reflected in the evaluation.  All relevant events (views, downloads, searches) have timestamps, making it possible to simulate what recommendations Obelix would have created, had it been running at the time the dataset was collected. Allowing stepping through time, creating recommendations given user at a given time, hide all selections for the current user and then attempt to recommend items for that user. However, since processing ten years of data is computational expensive, several approaches may be chosen to perform this task.  \\

Since the goal of the offline experiments is to filter algorithms and parameters, the test data sets should match as closely as possible to the real data the algorithm will handle in production. For that particular reason, it is important to assure that there is no bias in the distribution of users or items tested. It may be tempting to exclude users or items with low counts, in order to reduce the cost of the experiment \cite{microsoftevaluation}. However, when excluding users with some properties, a systematic bias is introduced and may lead to complicated corrections later when evaluation the results. A preferable approach may be to sample users and items randomly, considering only a subset of the data. However, the random sampling may introduce another bias towards favoring parameters and algorithms that work better with more sparse data \cite{microsoftevaluation}. In order to to minimize any biases that may occur in these experiments, all tests are simulated multiple times, with a new set of sampled data, in order to obtain the most accurate results possible.  \\

\subsection{Available procedures}

As described in the introduction to offline experiments (see section \ref{ranking_measures}), multiple approaches may be chosen based on the amount of data and properties available. The most extensive method simulates every user action and then tries to predict the next action for a given user, this is described in \cite{microsoftevaluation} as the most accurate method. However, again due to limited hardware available (two, single eight-core, 16 GB machines). This method is simply not possible to perform for a total of 40 000 users, 1,4 million items and about 250 million relevant user interactions. It is necessary to introduce a trade-off, and the most attractive compromise is to sample user interactions and time randomly. By randomly sample user interactions based on the entire data set, the goal is to minimize any potential bias that may be introduced by looking at a subset of the data. 

The most comprehensive method explained in \cite{microsoftevaluation} begins with no available prior data. Then steps through every user selection in temporal order, attempting to predict each selection and then making the choices available for future predictions. 

A simpler approach is to sample test users and time randomly just prior to a user action. Then hide all selections of all users after that moment, and then attempt to recommend items to that user. This approach requires changing the set of given information prior to each recommendation, which can still be computationally quite expensive. 

An even simpler alternative is to sample a set of test users, using a single test time, and then hide all items after the sampled test time for each test user. Resulting in simulating a situation where the recommendation system contains data as it would have at the moment of the test, and then makes recommendations without taking into account any new data that arrives after the test time.  

\subsection{Evaluating the current Invenio Search Engine}
In this experiment, the existing search engine in CDS was tested, in order to get an overview of how well the current search engine are able to satisfy the users needs.

\subsubsection{Process}
All the logs from the past ten years are indexed into an ElasticSearch cluster, making the logs available to query. ElasticsSearch supports to filter by terms and by timestamp ranges when searching.  

This experiment is designed to test the system before a new system is introduced, in order to gather data on the quality of the search results. In other words, how many items does the user has to scroll through, in order to find the desired item.  

The procedure was to create a tool to step through the logs, looking for patterns such as: \textit{user x searches with query y, and then clicked on item z}. However, this pattern is not enough to tell anything about the position of the item in the search result. For that reason, it was necessary to extend the tool to be able to re-do the search performed by \textit{user x}, with \textit{query y}, and then look for the position of \textit{item z} in the simulated search result. In order to test this as accurately as possible, only anonymous users was considered. The reason for using anonymous users is that those users have the same permissions, and will for that reason retrieve the same search result when searching with the same \textit{query y}. Anonymous users are users who are not logged in, and might be users who work at CERN, users form collaborating organizations or users who stumble upon CDS when browsing the web. 

However, even when using anonymous users with the same permissions, there is a chance for the search results to differ, the reason for this is that new items are introduced every day through new submissions or harvesting of items. In order to tackle this issue of different search results over time, only recent logs are used in this experiment. It also makes sense to use recent logs, in order to test the system as it is, not how it was several years ago. For the aforementioned reasons, the experiment was conducted on 1 day old logs and running for three weeks, collecting 21 days of data. When re-doing the searches, all the searches performed by the tool the day before is ignored, this is possible by leaving a mark in the url, which does not affect the search result, but makes it possible to identify that the search was done by the tool.

\subsubsection{Results}
The data collected during the 21 days long experiment resulted in a total of approximately 500 000 searches, with an average of 25 000 searches a day. This is too many searches data for a single test machine to re-do in a reasonable amount of time (and without overloading the CDS servers with a thousands of new queries). Instead of re-doing all searches, a random sample of searches for each day with a size of 500, meaning that 10 000 of the searches were reproduced. 

\begin{figure}
  \caption{Average click position in search results using latest first ranking}
  \label{fig:average_click_position_using_latest_first}
  \centering
    \includegraphics[width=1.0\textwidth]{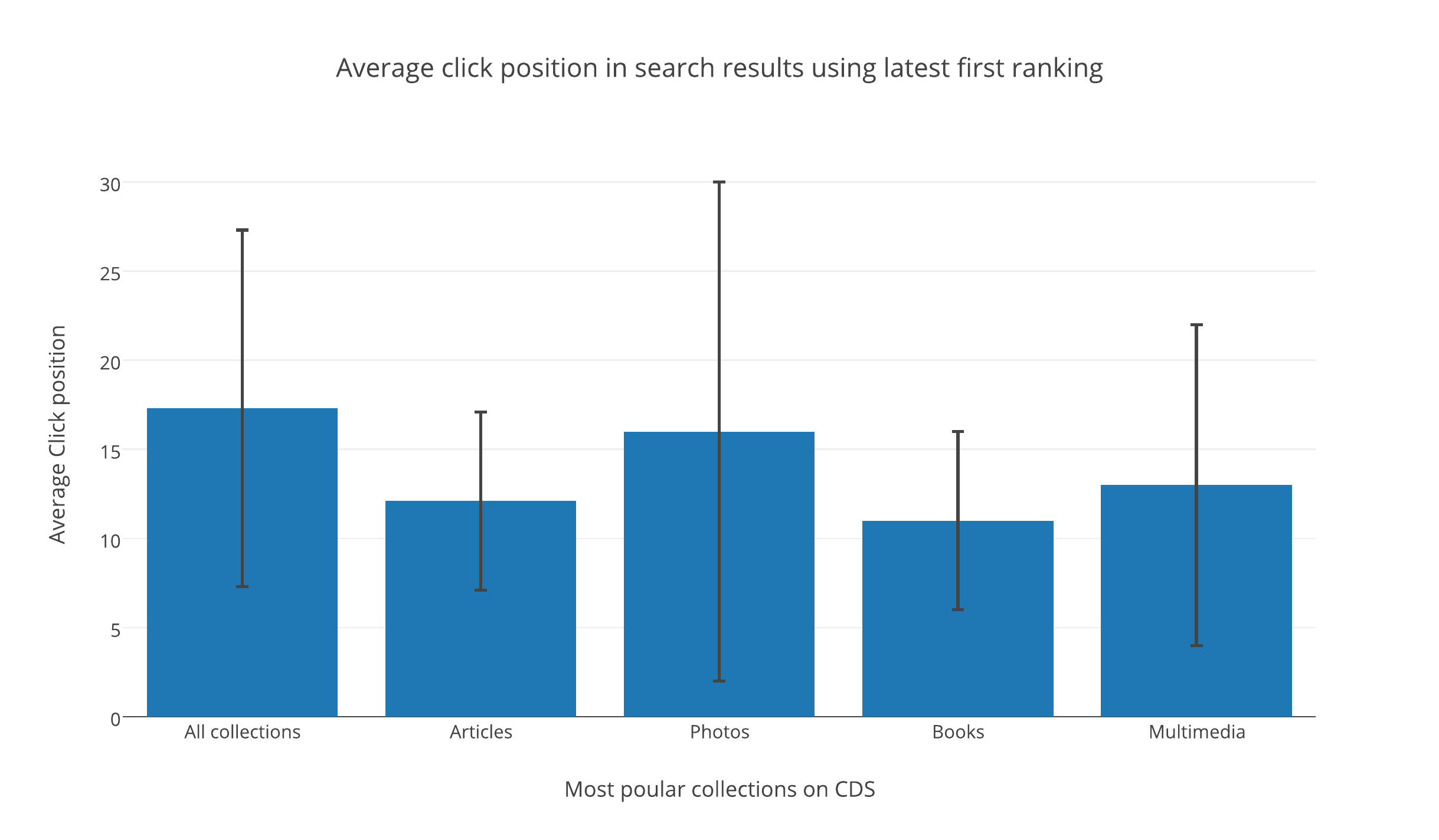}
\end{figure}

The result of this experiment is plotted in figure \ref{fig:average_click_position_using_latest_first}, and shows the average number of items the user has to scroll through in order to find the desired item. The global search covers when users search using the simple search from the home page of CDS, the other bars are the most popular collections to search, whereas the remaining collections are excluded from the result due to limited data.

The method used to rank the results is \textit{latest first} (default on CDS). The error bars indicate how large the variance is inside each collection. From the plot it is clear that about 50\% of the users needs to scroll through more than 10 items in order to find the desired item. This means that on CDS, about 50\% of the users have to click on the \textit{next page} link, in order to find the desired item.

An observation is that the searching within a collection makes it easier to find the desired item, this observations makes sense as the user already helped the search engine filter out irrelevant items by choosing a collection, in addition, a collection contains less items than \textit{all collections} combined, making \textit{latest first} ranking more useful within a collection.

Another observation is that the variance of photos and multimedia is higher than for articles and books. This also makes sense for two reasons, the first is that photos and videos lack extensive meta data as discovered in the user survey \cite{lemeur}, and that the users might not be interested in the latest photos or vidoes in the same way the users are interested in the latest publication within a subject. For instance, when a user search for a photo of the \textit{Large Hydron Collider} at CERN, the users might be more interested in finding the best photo, not necessarily the most recent one. 

\subsection{Evaluting prediction accuracy of Obelix}
\label{predicting_accuracy}
In order to personalize the search engine using a recommendation system, the predictions of the recommendation system have to be useful as described in section \ref{bg:prediction_accuracy}. This experiment explores how accurately the prototype of Obelix is able to predict the user behavior in CDS. In this experiment the goal is to test the proposed prototype of Obelix with different parameters and time ranges, in order to find the best possible combination to predict the users behavior.

\subsubsection{Process}
The approach for this experiment is based on the methods explained in 
\ref{sec:evaluation_offline_experiments}, inspired by the work by Microsoft on evaluation of recommendation systems  \cite{microsoftevaluation}. The approach chosen for this experiment is to sample a set of users' interactions for a time period, and then attempt to predict the next interactions for that user. This work includes dividing the collected logs into two data sets, a training set and a test set. The training set is used to build a graph, related users and their interactions with other users. Then the test is used to validate the user predictions by measuring the amount of correct predictions. The data sets consist of users' usages of items, split into two categories: downloads and views. In this experiment, the following parameters are tested: \\

\begin{description}

\item[Time frame: \textit{t}] \hfill \\
How much data is necessary to be able to predict, and is too much data (old data) limiting the accuracy. 
Tested values are one year, two years and three years.

\item[Number of usages included: \textit{n}] \hfill \\
Should all the usages by a user be included or should only a limited number of usages be included, for instance the latest 30 usages.
Tested values are: [25, 50, 75, 100, 125, 150, 175, 200]

\item[Graph traversal depth: \textit{d}] \hfill \\
How deep should the algorithm traverse the graph from a given user? If there are no limits, the entire graph might be included. The tests are limited to a depth of eight, the reason for this is that based on observations, most of the graph is covered before reaching a depth of 5.
Tested values are: [1, 2, 3, 4, 5, 6, 7, 8]

\item[Item usages weight: \textit{w}] \hfill \\
How important is each usage, if an item is used by twice as many users as another item on the same depth in the graph, is it twice as relevant for the current user? In order to test this, four different techniques are used to calculate the final score of an item: [constant factor 1, log, normalized, log-normalized]

\end{description}

\noindent{
The testing and evaluation of these parameters lead to the development of a new tool: \textit{Obelix Testing Framework}. This tool supports defining all these parameters, then searching through ElasticSearch to find matching item usages, and finally measure the prediction accuracy. The \textit{Obelix Testing Framework} produces csv-files with all parameters listed as columns, sorted by the best prediction accuracy. In order to make use of parallel execution and save time, the experiment was split into several runs with different time ranges. All other parameters were tested inside each time range. 
}

Three different training sets were used, while keeping the test set constant. The three different training sets represent one year, two years and three years of training. The test set is kept constant in order to compare the three different training sets. Within each test representing a training set, all other variables are tested with the testing values described in the list above. The \textit{Obelix Testing Framework} simulates the steps for a sampled set of 100 users, and the simulations are performed three times with random sampled users each time. The entire graph is built, including all users' interactions, but only the steps of the 100 sampled users are tested and including in the result. 

The choice of only testing three different training sets and no more than 100 sampled users is justified by the computing power available. A total of about six million different combinations were tested, and for each combination an average of 75 predictions per user was tested. Leaving us with a total of about 90 million predictions. Each prediction took about 15ms to perform, which means that the experiment ran for about 14 days. 

In order to test the predicted selections generated by Obelix, Obelix generated a list of items that a user \textit{should be interested in}, and compare this list to the actual selected items. For this test to be valid, the list of recommendations is limited to the number of selected items by the user (an average of 75 items for six months). This comparison will then tell us how many of the top 75 recommendations that was successful. 

This test does not provide a perfect overview of the prediction accuracy, but instead show trends of how the different variables affect the result.

\subsubsection{Results}
The result are split into the three different training sets, and the top 10 results is shown in figure x, y and z. Each parameter is plotted with regard to RMSE and the three data sets. 

\begin{figure}[!ht]
  \caption{Number of interactions impact}
  \centering
    \includegraphics[width=1.0\textwidth]{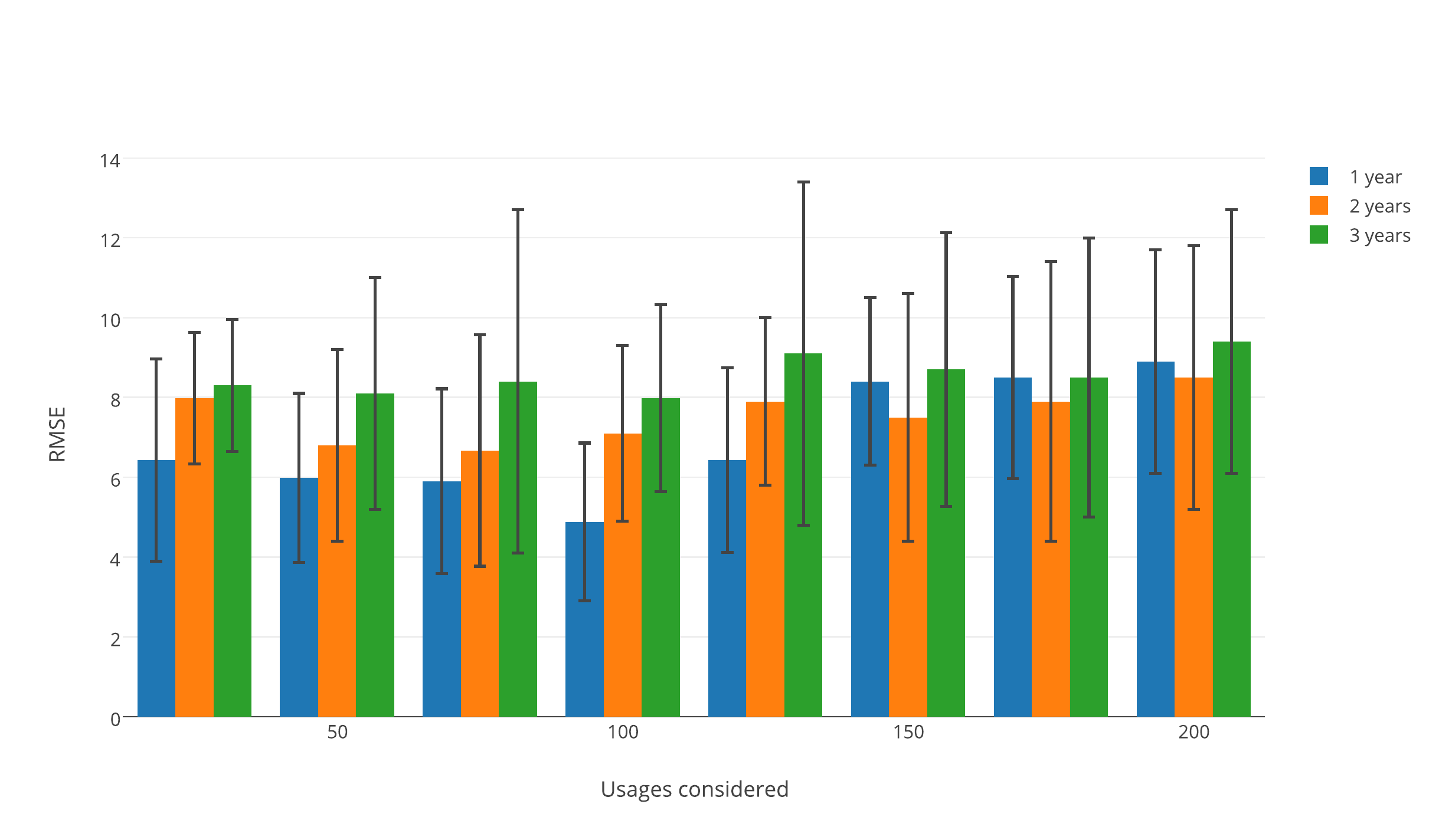}
  \label{fig:numberofinteractions-rmse}
\end{figure}

Starting with the usages considered (\textit{n}). This plot shows how the number of usages used in the graph affect the predictions. As seen in figure \ref{fig:numberofinteractions-rmse}, the different values [25,50,75,100,125,150,175,200] produces almost the same result. It is not clear from the results which one is the best fit. However, there are three key elements worth to noticing:

\begin{itemize}
\item The predictions are almost equally good with 25 usages as with 100. 
\item It is better to focus on the most recent usages, and not include several years of data.  
\item It is a trend for the performance of the predictions to flat out for 150 usages and more. 
\end{itemize}

\begin{figure}[!ht]
  \caption{Traversal depth impact}
  \centering
    \includegraphics[width=1.0\textwidth]{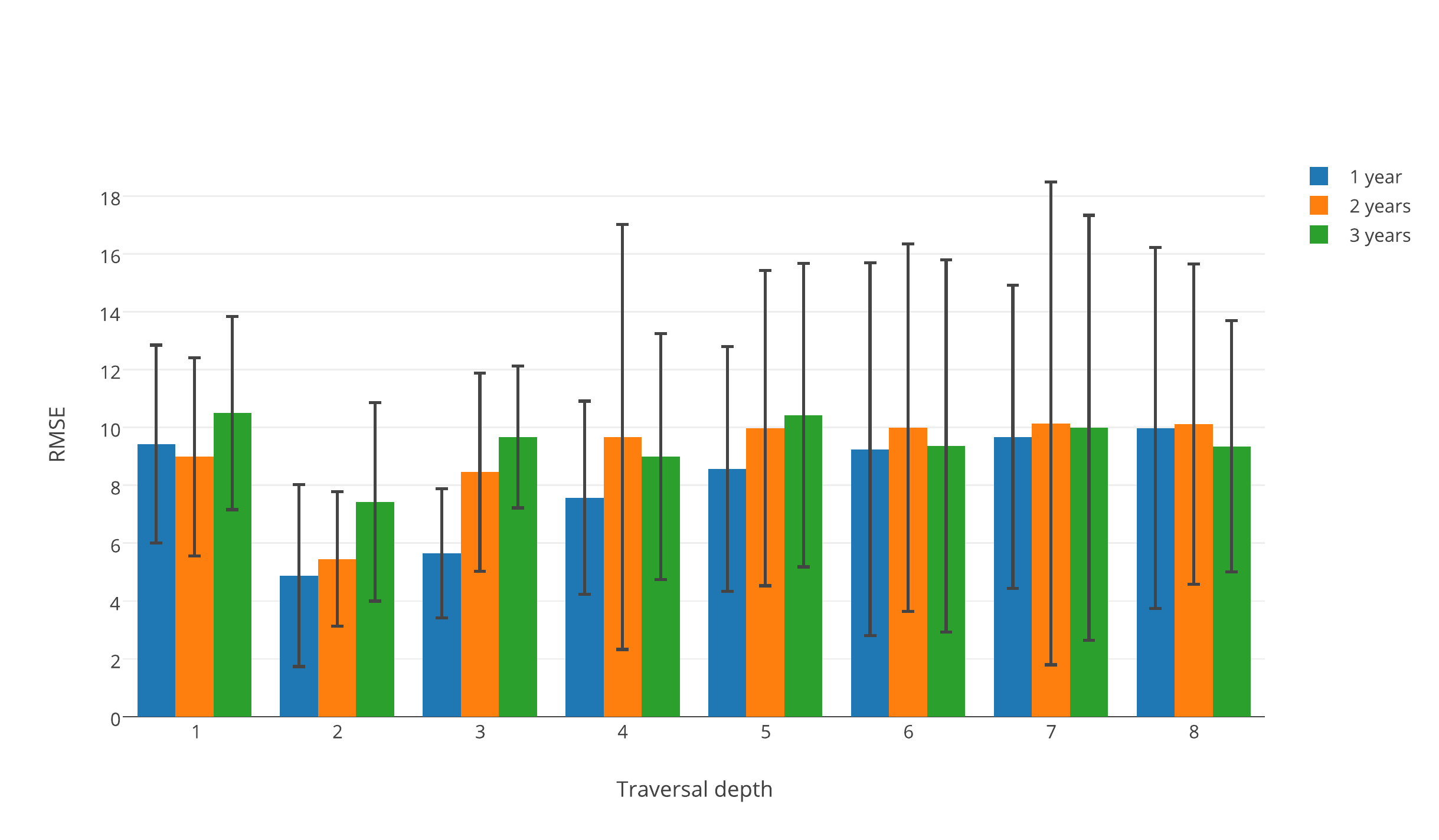}
  \label{fig:traversal-depth-rmse}
\end{figure}

The second test is for the Graph traversal depth \textit{d}. The plot in figure \ref{fig:traversal-depth-rmse} shows a clear gain for using 2 or 3 as the depth. Another aspect worth noticing is that the performance flat out from 5. The reason for this is most sub-graphs of the users, most items are reached between a depth from 5 to 8. 

This is as expected, the core idea is based on the assumption that similar users belong to a \textit{community}, or that they share the same interest in items. If the the graph is traversed to deep, all the users become a part of the community.

\begin{figure}[!ht]
  \caption{Weighting method impact}
  \centering
    \includegraphics[width=1.0\textwidth]{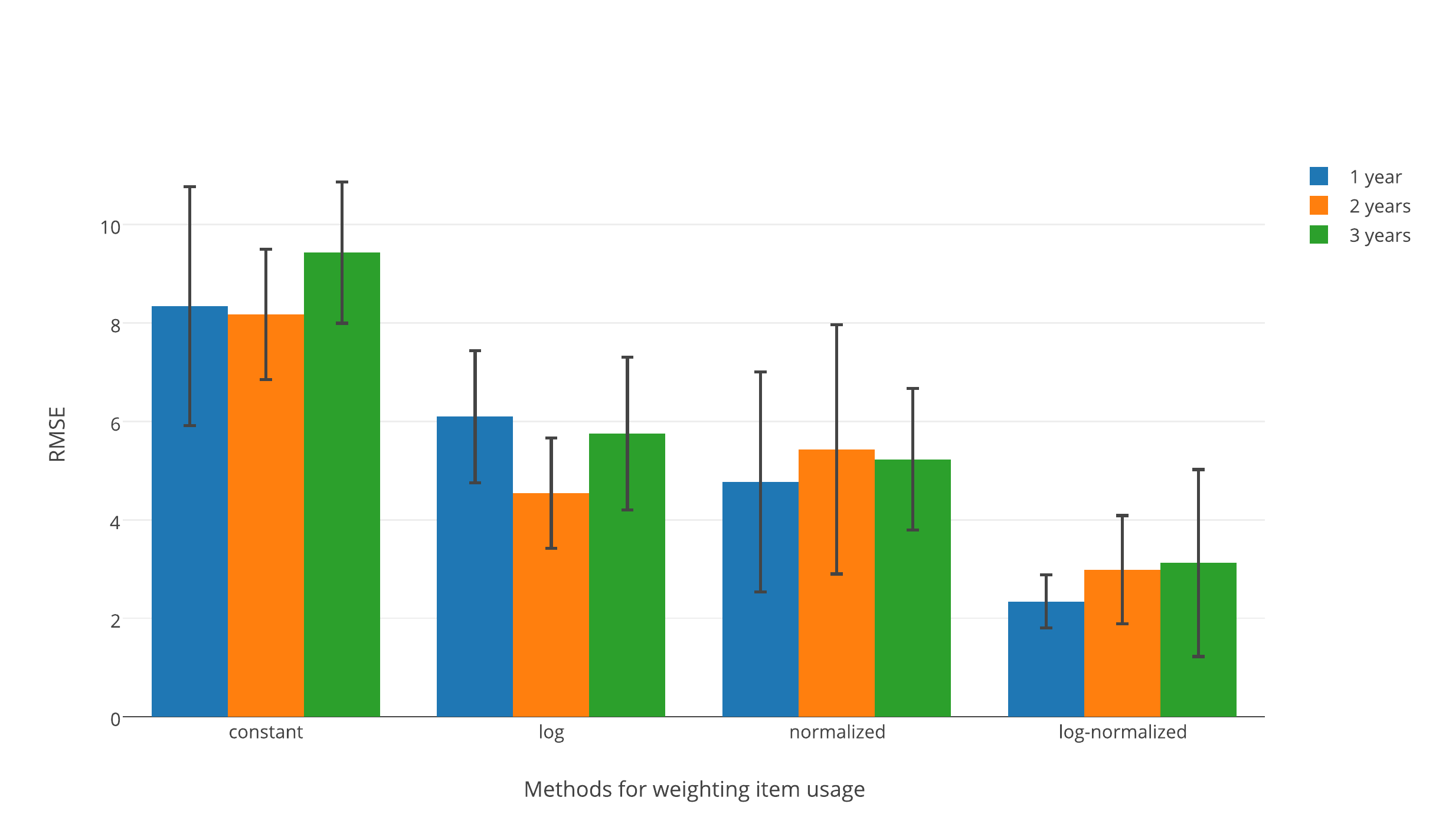}
  \label{fig:weighting-rmse-item-usage}
\end{figure}

\section{Online experiments} 

\subsection{Introduction}
In any real recommendation system, the goal is to influence how the user interact when searching. For this reason, it is interesting to measure to what degree the system changes the user behavior. For example, it is interesting to see to what extent the users follows the recommendations. What is even more interesting to measure is how many items the user has to consider (scroll through) in order to find the desired item. Ideally, the most relevant document should be the first (on top) of the result set or at least shown as part of the top n items in the result. On CDS, top-n is ten per collection by default. In order to measure the real effect of a recommendation system, a set of experiments needs to be executed. The interesting point being to compare Obelix ranking to other ranking methods such as word similarity and the default latest first. Ranking quality is typically evaluated using DCG (Discounted cumulative gain),  based on two assumptions: 

\begin{itemize}

  \item Highly relevant documents are more useful when appearing earlier in a search engine result list (have higher ranks)

  \item Highly relevant documents are more useful than marginally relevant documents, which are in turn more useful than irrelevant documents.
DCG originates from an earlier, more primitive, measure called Cumulative Gain.

\end{itemize}

These evaluation methods are carefully chosen based on advices from \cite{microsoftevaluation} and the data set available. All these methods contribute to the understanding the quality and usefulness of Obelix. For the evaluation to be valuable, a requirement is to compare it with the current available methods, in order to conclude and discuss the results and performance. 

\subsection{Evaluating Click Position}
A simple but efficient method to evaluate the performance is to observe  which position (scrolling records) in the search result the user find something relevant. For instance, if the user is looking for something new, it should be on top of the results when sorting by \textit{latest first}. However, when the user is looking for a popular item produced in the year 2000, it may be on the bottom when sorting by latest first and 1st when sorting by recommendation or word similarity. Obelix attempts to solve searching in general, meaning that the goal of Obelix is to rank better than both word similarity and latest first for most users and in most cases.

The goal is to compare Obelix with both \textbf{latest first} and \textit{word similarity}.

\subsubsection{Setup}
During the online experiment period of the two first months of 2015 on the CERN Document Server, a new default ranking method was introduced without informing the users. The new default ranking method "recommendations" was implemented with a dynamic settings file, enabling multivariate testing (see section \ref{multivariate_testing}) by constantly changing the settings for users. If the user wanted to, it was always possible to disable the recommendations by switching back to sort by latest first or word similarity. 

The dynamic settings file had one option of particular interest, the importance factor. Since Obelix implements a second ranking method, the importance factor enables a dynamic experiment environment. By setting the importance factor to zero, the recommendations provided by Obelix was ignored, while an importance factor of one totally overrides the first ranking method. During the experiment, the importance factor was dynamically switching between [0, 0.3, 0.5, 0.6, 0.9, 1] randomly every tenth minute. Providing a controlled environment where the performance impact of Obelix is observable and possible to measure. Figure \ref{fig:click_position_over_time} shows a comparison between the importance factors, "latest first" and "word similarity from the online experiment. \\

The data was collected in two phases during the test period. The first phase included logging searches and views by the users; the second was to re-do all searches done by the users with the \textit{word similarity} and \textit{latest first} ranking. By re-doing the searches and scroll through the results to find the item the user viewed originally allowed the comparison of the three methods (\textit{latest first}, \textit{word similarity} and \textit{Obelix}).  

Unfortunately, due to restriction of items in CDS, and a continuous growing set of items on CDS, some issues arise with this approach. 

\begin{itemize}

  \item Restricted items are impossible to find for anonymous users, making it necessary to skip those searches and views, and thereby end up with a smaller test set. (The evaluation shows that it is needed to skip about 30\% of the searches). 

  \item Some users might find restricted items while others do not, the result contains items based on the user's permissions, making the re-do of searches using an anonymous user inaccurate. (It is important to notice that different users have access to a different set of restricted items). 

  \item New items are added to the collections continuously, new items may be a part of the result of the latter searches, again making the search inaccurate. 

  \item The approach of re-do searches with different ranking methods introduces an unwanted bias towards good results for recommendations. The measurement used is how many items the user has to scroll through before clicking, assuming that the user only click and view what is relevant. However, users presented with a different result might select different items to click and view.
  
\end{itemize}

\subsubsection{Result}

\begin{figure}[h!]
  \caption{Ranking methods comparison}
  \centering
    \includegraphics[width=1.0\textwidth]{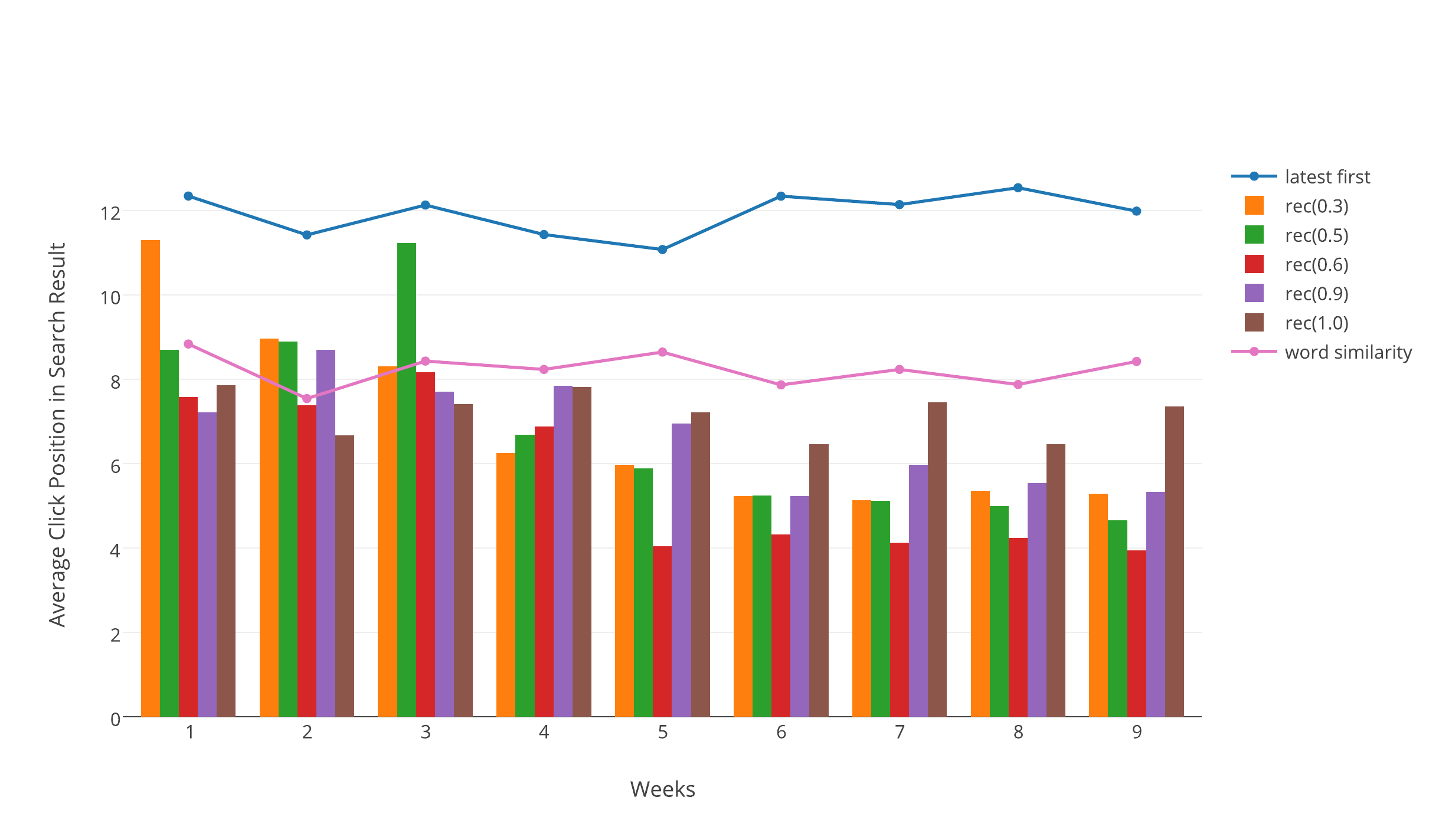}
   \label{fig:click_position_over_time}
\end{figure}

As seen in figure \ref{fig:click_position_over_time}, the search results denoted as rec[0,1] outperform both latest first and word similarity for global search. Another observation is that the click position has improved during the online experiment, whereas both latest first and word similarity remain stable.

\subsection{Evaluating effectiveness search time} 

In order to meet the requirements from CDS on performance, it is important for measure the effectivenes. For that reason, the average time to render the search results are plotted for both rrm (recommendations), wrd (word similarity) and ltf (latest first).

The interesting observation from this plot is that the performance of Obelix is decreasing, where as the latest first is increasing. This behavior can be explained, the amount of data transfered and processed by Obelix has increased. Regarding latest first, the reason for the increase in performance is that only users who specifically has selected to use ltf uses in the last 4 weeks, becase using Obelix is set as default. 

\begin{figure}
  \caption{Seconds to load the search result with using different ranking methods}
  \label{fig:seconds_to_load_search_result}
  \centering
    \includegraphics[width=1.0\textwidth]{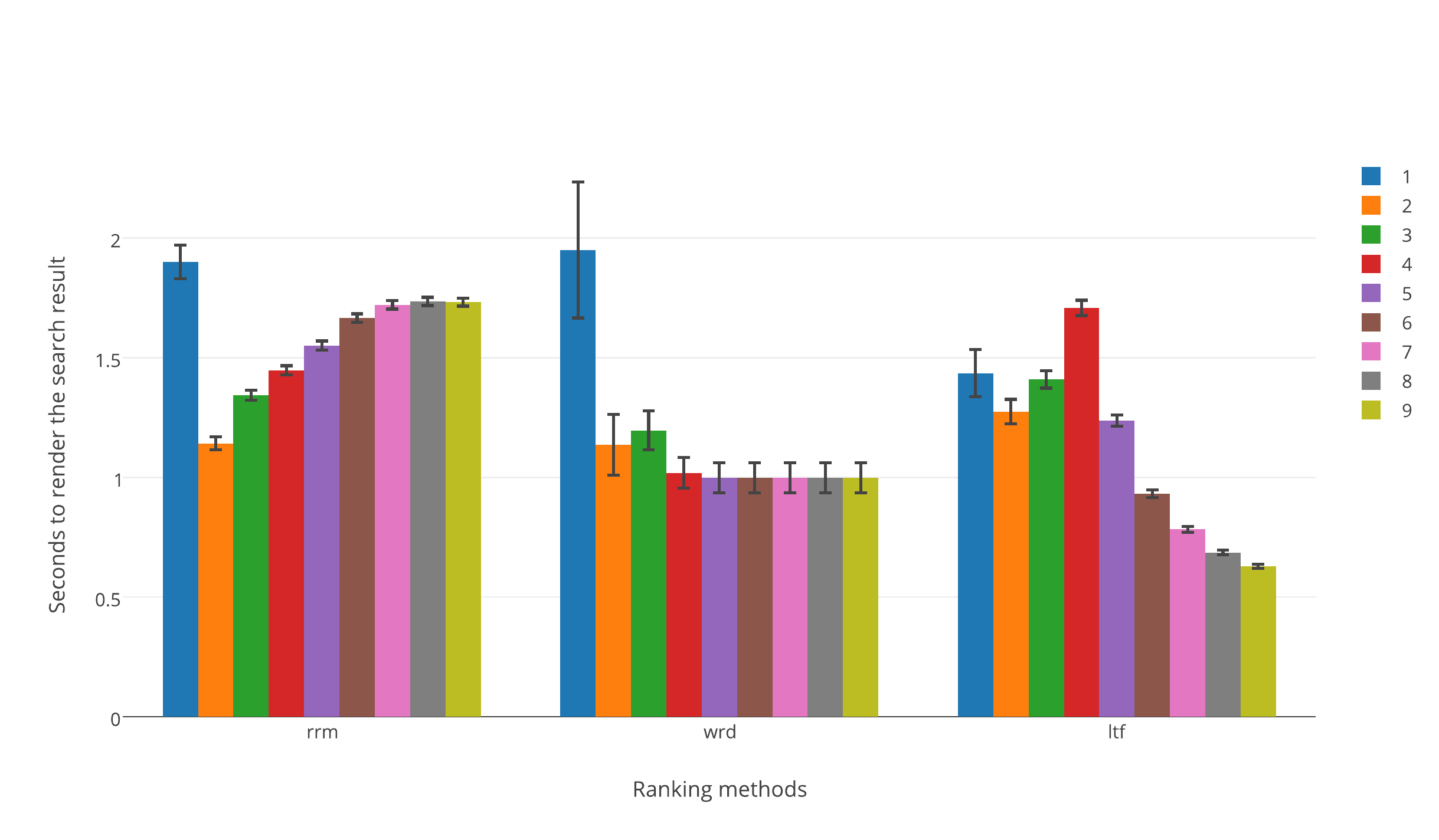}
\end{figure}

\chapter{Conclusion}
This chapter includes a summary of the accomplishments and contributions to the community of search and Invenio. This chapter also evaluates the project as a whole, and proposes future work. 

\section{Summary}
This work was motivated by the need for a better search engine on the CERN Document Server (CDS). CDS is powered by Invenio, an open source digital library system developed at CERN. In order to improve the search engine, a new recommendation system used to personalize the search results was developed. 

The need for a better search engine was based on user surveys, interviews with users and librarians, related work on search in Invenio, as well as an experiment conducted to evaluate the efficiency of the current search engine. 

The feedback and related work surveyed for this project led to the development of Obelix, a new recommendation system, built to serve the purpose of providing personalized search. In the process of planning Obelix, a set of requirements was agreed on with CERN in order to allow Obelix to run in production. 

The prototype of Obelix was evaluated using offline experiments in order to find the best possible combination of parameters. The offline experiments showed that is was possible to predict about 45\% of the user interactions with only a year's worth of collected data. This was found promising and led to the development of a fully functional implementation of Obelix, which was run in production.

The online experiments running in production for two months collected enough data to show that the personalized search experience outperformed existing ranking methods when performing global search. In \textit{latest first} and \textit{word similarity}, which are the previously used methods, the average click position is 12.12 and 9.34, respectively. When using Obelix, the average click position is 5.64, which is clearly an improvement compared to the existing ranking methods. However, this is not true when searching collections individually, for instance, the results from the \textit{CERN Yellow Reports} collection showed that \textit{latest first} is the preferred ranking method, and for the \textit{books collection}, the \textit{word similarity} produced the best results.

Based on the results from the online experiments, it is clear that the new personalized search engine can not replace \textit{latest first} and \textit{word similarity} in all use cases, but it is certainly an improvement for the global search.

Consequently, it is necessary to evaluate each collection separately, in order to decide the default ranking method for that specific collection. However, based on results produced by the online experiments, it is recommended to keep the personalized search enabled as default for the global search. 

\subsection{Contributions}
This section lists the contributions from this project. 

\begin{description}

\item[Obelix] \hfill \\
The new recommendation system named Obelix is developed from scratch. 
This work includes designing and implementing the algorithm used, the REST API, as well as importing new entries. 

This work also includes swappable backends for queues and key/value-stores, as well as the graph database. Every backend-technology is swappable, allowing the integrator to write their own integration with a different graph database or queue.

\item[Obelix Caching Framework] \hfill \\
This module maintains the caching of Obelix, using queues and key-value stores. This work was done to improve the performance of Obelix, and to match the requirements from CDS. The CDS implementation makes heavily use of Redis, but the backends are swappable.  

\item[Obelix Invenio Integration] \hfill \\
Obelix is integrated into Invenio as an optional module, making it possible to enable/disable the recommendation feature at any time.

\item[Obelix Configuration Distribution] \hfill \\
As the the integration with Obelix require some parameters, it is useful to make Obelix manage the configuration for the clients. This is distributed using Redis.

\item[Obelix Logging Framework] \hfill \\
In order to collect data to generate useful statistics, a logging framework was developed to log and aggregate statistics.

This led to the development of an ELK-cluster dedicated to store statistics from Obelix. 

\item[Obelix Graph Visualizer] \hfill \\
The need for visualizing the underlying graph for Obelix led to the development of a dedicated visualization tool. This tool is built using WebGL, allowing the browser to handle thousands of nodes and edges simultaneously.

\end{description}

\section{Evaluation}

The experimental results show that the ranking of search results are possible to improve. However, even though the results are not exceptional, they show a high potential. The motivation for this project was to improve the ranking of search results when using the CERN Document Server. Obelix has demonstrated the ability to learn the users' interests and recommend the relevant items. The ability is evident from the results of the experiments; the average number of items the user has to scroll through is smaller with Obelix enabled. The improved search is not only a verification that Obelix works but also an argument for personalizing search results. 

While previous work on improving the search engine of CDS attempts so solve all challenges with search, Obelix focuses specifically on trying to find hidden treasures. Items that users want to find, but because of bad ranking, they would have to scroll through several pages to find. The core idea of Obelix is that is should only rank, not introduce or remove any items from the original result set. This idea was essential in order to implement it on an existing service like CDS. Previous attempts at improving the search by recommendations have ended in introducing items that might be relevant to the user but not the query. For that reason, Obelix was developed with the goal of ranking, not recommending items on a general basis.

\section{Future Work}

Future work includes improving the search for collections that are best suited for latest first or word similarity today. This should be possible as as these items share the same property of being popular within communities. 

Another feature work would be to include a notion of trust in the collaborative filtering algorithm, research has pointed out a trend to rely more on recommendations from friends and other people they trust than recommendations from anonymous similar users (see section \ref{background:trustenhanced}). 


\renewcommand*{\bibname}{References}
\bibliographystyle{alpha}
\bibliography{main}

 \appendix
 \addtocontents{toc}{%
  \protect\vspace{1em}%
  \protect\noindent \bfseries \appendixtocname\protect\par
  \protect\vspace{-.5em}%
 }
 \renewcommand{\chaptername}{\appendixname}

\chapter{Invenio modules}
\label{App:AppendixA}
\begin{itemize}
    \item \textbf{BibCheck} permits administrators and library cataloguers to automate various kind of tests on the metadata to see whether the metadata comply with quality standards. For example, that certain metadata fields are of a certain length, that they have numeric content, that they must not be present when other field exists, that their content is governed by an authority base depending on values of other fields, etc. The module can report its findings or can even automatically correct some kind of errors.
\item \textbf{BibClassify} allows automatic extraction of keywords from fulltext documents, based on the frequency of specific terms, taken from a controlled vocabulary. Controlled vocabularies can be expressed as simple text thesauri or as structured, RDF-compliant, taxonomies, to allow a semantic classification.
\item \textbf{BibConvert} allows metadata conversion from any structured or semi-structured proprietary format into any other format, typically the MARC XML that is natively used in Invenio. Nevertheless the input and output formats are fully configurable and have been tested on data importations from more than one hundred data sources. The power of this utility lies in the fact that no structural attributes of data source are presumed, but they are defined in an extensive data source configuration. Inevitably, this leads to a high complexity of the BibConvert configuration language. Most frequent configurations are provided with the Invenio distribution, such as a sample configuration from Qualified Dublin Core into the MARCXML. 
In general the BibConvert configuration consists from the source data descriptions and target data descriptions. The processor then analyzes and parses the input data and creates the resulting data structure, similarly as the XSLT processor would do. Typically the BibConvert is aimed at usage for input data that do not dispose of an XML representation. The source data is required to be structured or semi-structured, (i.e. not expressed in natural language that is a subject of information extraction task) and its processing involves several steps including record separation and field extraction upto transformation of source field values and their formatting.
\item \textbf{BibEdit} permits one to edit the metadata via a Web interface.
\item \textbf{BibFormat} is in charge of formatting the bibliographic metadata in numerous ways. This truly enables the separation of data content administration and formatting layout. BibFormat can act in the background and format the records when needed, or can preformat records for some often used outputs, such as the brief format used when displaying search results.
The BibFormat settings can be administered either through a user-friendly web interface, or directly by editing human-readable configuration files.
\item \textbf{OAIHarvest} represents the OAi-PMH compatible harvester allowing the repository to gather metadata from fellow OAi-compliant repositories and the OAi-PMH repository management. Repository is built directly on top of the database and disposes of an OAi repository manager that allows to perform the administrative tasks on the repository aside from the principal generic data administration module. The database can be partially or completely open for harvesting in the scope of the OAi-PMH protocol. In this case, all data is provided in raw form, where the semantics of individual tags is indicated uniquely by the MARC21 naming convention. This is particularly interesting for institutes that are specialized in cross-archive and cross-disciplinary services provision, as for example the ARC service provider.
\item \textbf{BibIndex} module takes care of the indexation of metadata, references and full text files.  Two kinds of indexes -- word and phrase index -- are being maintained.  The user can define several logical indexes (e.g. author index, title index, etc.) and the correspondence of which physical MARC21 metadata tag goes into which logical field index.  An index consists of two parts: (i) a forward index listing various words (or phrases) found in the given field, with the set of record identifiers where the given word can be found; and (ii) a reverse index listing record identifiers, with the set of words of the given record that go to the forward index.  Such a two-part indexing technique allows one to rapidly update only those words that have changed in the input metadata record.  The indexes were designed with the aim to provide fast user-response search times and are faster than native MySQL (full text) indexes.
\item \textbf{BibMatch} permits to filter input XML files against the database content, attempting to match records via certain criteria, for example to avoid doubly-inputted records.
\item \textbf{BibRank} permits to set up various ranking criteria that will be used later by the search engine. For example, ranking by the word frequency, or by some metadata tag value such as journal name by means of the journal impact factor knowledge base, or even by the number of downloads of a particular paper. Note that BibRank is independent of BibIndex.
\item \textbf{BibSched} The bibliographic task scheduler is central unit of the system that allows all other modules to access the bibliographic database in a controlled manner, preventing sharing violation threats and assuring the coherent execution of the database update tasks. The module comes with an administrative interface that allows to monitor the task queue including various possibilities of a manual intervention, for example to re-schedule queued tasks, change the task order, etc.
\item \textbf{BibUpload} allows to load the new bibliographic data into the database. To effectuate this task the data must be a well-formed XML file that complies with the current metadata tag selection schema. Usually, the properly structured input files of BibUpload come from the BibConvert utility.
\item \textbf{ElmSubmit} is an email submission gateway that permits for automatic document uploads from trusted sources via email. (Usually web submission or harvesting is preferred.)
\item \textbf{MiscUtil} is a collection of miscellaneous utilities that other modules are using, like the international messages, etc.
\item \textbf{WebAccess} module is responsible for granting access to users for performing various actions within the system.  A Role-Based Access Control (RBAC) technique is used, where users belong to several groups according to their role in the system.  Each user group can be granted to perform certain actions depending on possible one more action arguments.  WebAccess is presently used mainly for the administrative interface.  There are basically two kinds of actions: (i) configuration of administrative modules and (ii) running administrative tasks.
\item \textbf{WebAlert} module allows the end user to be alerted whenever a new document matching her personal criteria is inserted into the database.  The criteria correspond to a typical user query as if it would be done via the search interface.  For example, a user may want to get notified whenever a new document containing certain words, or of a certain subject, is inserted.  A user may create several alerts with a daily, weekly, or a monthly frequency.  The results of alert searches are either sent back to the user by email or can also be stored into her baskets.
\item \textbf{WebBasket} module enables the end user of the system to store the documents she is interested in in a personal basket or a personal shelf.  The concept is similar to popular shopping carts.  One user may own several baskets.  A basket can be either private or public, allowing a simple document sharing mechanism within a group.
\item \textbf{WebComment} provides a community-oriented tool to rank documents by the readers or to share comments on the documents by the readers. Integrated with the group-aware WebBasket, WebGroup, WebMessage tools, WebComment is at the heart of the social network features of the Invenio software.
\item \textbf{WebHelp} presents some global user-level, admin-level, and hacker-level documenation on Invenio. The module-specific documentation is included within each particular module.
\item \textbf{WebMessage} permits the communication between (possibly anonymous) end users via web message boards, to invite readers to join the groups, etc.
\item \textbf{WebSearch} module handles user requests to search for a certain words or phrases in the database.  Two types of searching can be performed: a word search or a phrase search.  The system allows for complex boolean queries, regular expression searching, or a combined metadata, references and full text file searching in one go.  Users have a possibility to browse for present index terms.  If no direct match could have been found with the user-typed query pattern, the system proposes alternative matches as a search guidance.  The search indexes were designed to provide fast response times for middle-sized data collections of up to 106 records. 
The metadata corpus is organized into metadata collections that are directly accessible through the browse function, similarly to the popular concept of Web Directories.  Orthogonal views on the document corpus are enabled in the search interface via a concept of virtual collections: for example, a document may be classified both according to its type (e.g. preprint, book) and according to its Dewey decimal classification number.  Such a flexible organization views allows for the creation of easy navigation schemata to the end users.
\item \textbf{WebSession} is a session and user management module that permits to differentiate between users. Useful for personalization of the interface and services like personal baskets and alerts.
\item \textbf{WebStat} is a configurable system that permits to gather statistics about the health of the server, the usage of the system, as well as about some particular system features.
\item \textbf{WebStyle} is a library of design-related modules that defines look and feel of Invenio pages.
\item \textbf{WebSubmit} is a comprehensive submission system allowing authorized individuals (authors, secretaries and repository maintenance staff) to submit individual documents into the system. The submission system disposes of a flow-control mechanism that assures the data approval by authorized units. In total there are several different exploitable submission schemas at a disposal, including an automated full text document conversion from various textual and image formats. This module also disposes of information extraction functionality, focusing on bibliographic entities such as references, authors, keywords or other implicit metadata.
\end{itemize}

\end{document}